\documentclass[aps,prd,superscriptaddress,nofootinbib,floats,preprint,floatfix]{revtex4-1}
\usepackage{bm}
\usepackage{indentfirst}
\usepackage{amsmath}
\usepackage{graphicx}
\usepackage{float}
\usepackage{amssymb}
\usepackage{subfigure}
\usepackage{psfrag}
\usepackage{hyperref}
\usepackage{tensor}
\usepackage{braket}
\usepackage{multirow}
\usepackage{verbatim}
\usepackage{enumerate}
\usepackage{mathtools}
\usepackage{ulem}
\usepackage{cancel}
\usepackage{xcolor}
\hypersetup{
	colorlinks=true,
	linkcolor=red,
	citecolor=blue,
}
\allowdisplaybreaks[1]

 \newcommand{\bq}{\begin{equation}}
 \newcommand{\eq}{\end{equation}}
 \newcommand{\bqn}{\begin{eqnarray}}
 \newcommand{\eqn}{\end{eqnarray}}

\definecolor{XGDarkRed}{RGB}{200,0,0}

\begin{document}

\title{Scalar-induced gravitational waves in spatially covariant gravity with two tensorial  degrees of freedom}

\author{Jia-Xi Feng}
\email{fengjx@ucas.ac.cn}
\affiliation{School of Fundamental Physics and Mathematical Sciences, Hangzhou Institute for Advanced Study, UCAS, Hangzhou 310024, China}
\author{Yang Yu}
\email{yuy@tju.edu.cn}
\affiliation{Center for Joint Quantum Studies and Department of Physics, School of Science, Tianjin University, Tianjin 300350, China}
\author{Yan-Heng Yu }
\email{yhyu@ucas.ac.cn}
\affiliation{School of Fundamental Physics and Mathematical Sciences, Hangzhou Institute for Advanced Study, UCAS, Hangzhou 310024, China}
\affiliation{School of Physical Sciences, University of Chinese Academy of Sciences, Beijing 100049, China}
\author{Xian Gao}
\email{Corresponding author: gaoxian@mail.sysu.edu.cn}
\affiliation{School of Physics, Sun Yat-sen University, Guangzhou 510275, China}
\author{Qing-Guo Huang}
\email{Corresponding author: huangqg@itp.ac.cn}
\affiliation{School of Fundamental Physics and Mathematical Sciences, Hangzhou Institute for Advanced Study, UCAS, Hangzhou 310024, China}
\affiliation{School of Physical Sciences, University of Chinese Academy of Sciences, Beijing 100049, China}
\affiliation{Institute of Theoretical Physics,
Chinese Academy of Sciences, Beijing 100190, China}

\begin{abstract}
We study scalar-induced gravitational waves (SIGWs)
in a spatially
covariant gravity model with
two tensorial degrees of freedom, minimally coupled to a
$k$-essence scalar field. For the parameter choice considered during radiation domination, the quadratic
tensor action coincides with that of general relativity (GR).
We derive the scalar transfer functions at finite
$\sigma\equiv C_2\bar N$ and in the large-$|\sigma|$ limit.
We then construct the source function of SIGWs. 
We further examine a cubic extrinsic-curvature interaction that modifies the source function of SIGWs without affecting the
background or quadratic  actions.
We calculate the fractional energy density of SIGWs for monochromatic
and log-normal primordial curvature power spectra.
We find that,  for the positive values of $\sigma$ studied, finite-$\sigma$
effects suppress the spectrum
over a broad range of scales relative to
the result in the  $\sigma \rightarrow \infty$ limit. In this limit,  the cubic interaction suppresses the spectrum
at small wavenumbers and enhances it at larger
wavenumbers.
These results show that modified scalar dynamics
and non-linear interactions can alter the 
fractional energy density of SIGWs
even when the quadratic tensor action is identical
to that of GR.
\end{abstract}

\maketitle

\section{Introduction}

General relativity (GR) provides the standard description of
gravitational phenomena over a wide range of scales and has
successfully passed numerous observational tests.
Nevertheless, several open questions in cosmology, including
the origin of cosmic acceleration and the tension between
early- and late-time determinations of the Hubble constant,
continue to motivate the exploration of physics beyond GR
\cite{Perlmutter:1998np,Riess:1998cb,Verde:2019ivm,
DiValentino:2021izs}.
Modified gravity theories provide one possible framework for
addressing these questions and may leave observable signatures in cosmology \cite{Joyce:2014kja}.

The direct detection of gravitational waves (GWs)
has opened a new window on gravitational physics
\cite{LIGOScientific:2016aoc,LIGOScientific:2017vwq}.
A stochastic GW background of cosmological origin
could further provide information about the early
Universe. Among its possible sources,
scalar-induced gravitational waves (SIGWs),
generated by scalar perturbations at second order,
have attracted considerable attention
\cite{Ananda:2006af,Baumann:2007zm,Saito:2008jc,
Kohri:2018awv,Espinosa:2018eve,Inomata:2025wiv,Terada:2025cto}. In particular, the generation of SIGWs accompanying
primordial black hole (PBH) formation further
highlights their significance \cite{Saito:2008jc,Yuan:2019udt,Yi:2022ymw,
Chen:2021nxo,Kozaczuk:2021wcl,
Yi:2022anu,Cang:2022jyc,Zhao:2022kvz,Chang:2022nzu,Ghoshal:2023wri,Bi:2026zlt}.
Enhanced scalar perturbations on small scales,
which can generate sizeable SIGWs, can arise
in various inflationary models
\cite{Lin:2020goi,Fumagalli:2020nvq,Zhang:2021rqs,
Meng:2022low,Gu:2023mmd,Chen:2026scy}. The effects of primordial non-Gaussianity on
SIGWs have also been extensively studied
\cite{Cai:2018dig,Yuan:2020iwf,Adshead:2021hnm,
Zhang:2021vak,Chang:2023aba,Liu:2023ymk}. 
The anisotropies of induced GW backgrounds
\cite{Yu:2023jrs,Luo:2025lgr,Wu:2026jnh}, as well as
higher-order corrections and dissipative effects
on SIGWs
\cite{Chang:2023vjk,Zhou:2024ncc,
Yu:2024xmz,Yu:2025cqu},
 have also been investigated. For reviews, see
Refs.~\cite{Domenech:2021ztg,Yuan:2021qgz}.
The nanohertz signals reported by pulsar timing
array collaborations
\cite{NANOGrav:2023gor,NANOGrav:2023hde,
Zic:2023gta,EPTA:2023sfo,EPTA:2023fyk,Xu:2023wog}
have further stimulated
interest in SIGWs as a possible explanation
\cite{Chen:2019xse,NANOGrav:2023hvm,EPTA:2023xxk,Cai:2023dls,
Yi:2023mbm,Yi:2023tdk,Balaji:2023ehk,Wu:2023hsa,Liu:2023hpw,Wu:2024qdb,Chen:2024fir,Wu:2025gwt}.
Future space-based GW observatories, including
LISA, Taiji, and TianQin, will provide further
opportunities to probe stochastic GW backgrounds
\cite{LISA:2017pwj,Ruan:2018tsw,TianQin:2015yph,Jiang:2026krg}.

Since SIGWs are generated through non-linear mode coupling, they can  probe  gravitational interactions beyond the linear perturbation level.
In modified gravity, the evolution of scalar and tensor perturbations and their non-linear couplings can differ from those in GR, leading to distinctive features in the SIGW background. In particular, interaction terms that do not affect  the
background evolution or linear perturbation dynamics may nevertheless
enter the source of SIGWs.
SIGWs have been studied in a variety of modified-gravity
theories, 
including Brans--Dicke gravity \cite{Yi:2022anu}, $f(R)$ gravity \cite{Papanikolaou:2021uhe,Zhou:2024doz,Lopez:2025gfu},
Gauss--Bonnet gravity \cite{Zhang:2021rqs}, modified teleparallel gravity \cite{Papanikolaou:2022hkg,Tzerefos:2023mpe},
 Horndeski gravity
\cite{Domenech:2024drm}, 
and  GLPV gravity \cite{Domenech:2025zvi}.
SIGWs in parity-violating gravity have also been studied extensively, with parity violation potentially giving rise to circular polarization in the SIGW background
\cite{Zhang:2022xmm,Feng:2023veu,Zhang:2023scq,Zhang:2024vfw,Zhang:2025mps,Feng:2024yic,Feng:2026pcs}.
SIGWs have also been investigated in spatially covariant gravity (SCG) \cite{Jiang:2025ysb}. By preserving spatial diffeomorphism invariance while relaxing full spacetime diffeomorphism invariance \cite{Gao:2014fra,Gao:2014soa},  SCG admits a broad class of gravitational operators, with the resulting constraint structure determining the propagating degrees of freedom.

Current GW observations are consistent with tensor polarizations, with no evidence for additional polarization modes \cite{LIGOScientific:2017ycc,Takeda:2020tjj,Takeda:2021hgo,LIGOScientific:2021sio,LIGOScientific:2026qni}, motivating interest in theories that retain only the two tensorial degrees of freedom (DOFs) of GR. From a field-theoretic perspective, consistent local, Lorentz-invariant self-interactions of a single massless spin-2 field lead to GR under the restriction to at most second-order derivatives. Constructing alternative 2-DOF theories therefore requires relaxing some of these conditions. Among the modified theories discussed above, SCG provides a natural setting for this purpose \cite{Gao:2014fra,Gao:2014soa}. Although a class of SCG theories without higher order temporal derivatives naturally  propagates an additional scalar mode, appropriate conditions on the constraint structure can eliminate this mode, yielding theories propagating two tensorial DOFs \cite{Gao:2019twq,Lin:2020nro,Yao:2020tur,Yao:2023qjd}. Representative 2-DOF constructions include the cuscuton \cite{Afshordi:2006ad} and its extensions \cite{Iyonaga:2018vnu,Iyonaga:2020bmm}, as well as minimally modified gravity \cite{Lin:2017oow,Aoki:2018zcv,Aoki:2018brq,DeFelice:2020eju}. The cosmological implications of related 2-DOF theories have been extensively investigated in Refs.~\cite{Carballo-Rubio:2018czn,Aoki:2020oqc,DeFelice:2020cpt,DeFelice:2020onz,DeFelice:2020prd,DeFelice:2020ecp,Sangtawee:2021mhz,Aoki:2021zuy,DeFelice:2021xps,Pookkillath:2021gdp,Ganz:2022iiv,DeFelice:2022uxv,Jalali:2023wqh,Saito:2023bhn,Chakraborty:2023jek,Akarsu:2024qsi,Wang:2024hfd,Akarsu:2024eoo,Arora:2025msq,Ladeira:2026jne,Sangtawee:2026nym}. In particular, cosmological perturbations of 2-DOF theories coupled to an additional scalar matter field have been studied in Refs.~\cite{Iyonaga:2021yfv,Bartolo:2021wpt}.
Eliminating the gravitational scalar mode is intended to retain the same number of gravitational DOFs  as GR, while  the scalar matter field  is introduced to describe cosmological matter.
Perturbations of this matter field provide a source for SIGWs, allowing
us to explore the effects of modified gravity on their production.
Recently, explicit polynomial-type SCG models satisfying perturbative conditions required to eliminate the gravitational scalar mode through cubic order around a cosmological background were obtained in Refs.~\cite{Hu:2021yaq,Yu:2026fgn}.

In this work, we investigate SIGWs in a polynomial-type 2-DOF SCG theory \cite{Yu:2026fgn} minimally coupled to a $k$-essence scalar field with Lagrangian $P(\chi,X)$.
For general coefficient functions within the model, 
we derive the quadratic perturbation actions and
show that the reduced scalar kinetic matrix is
degenerate.
We then specialize to a  radiation-dominated
background with $P(\chi,X)=\lambda_0X^2$ and a parameter
choice for which the quadratic tensor action
coincides with that of GR.
We obtain the scalar transfer functions at finite
$\sigma\equiv C_2\bar N$ and in the large-$|\sigma|$
limit, and construct the SIGW source function and kernel.
We further examine a cubic interaction constructed from the traceless
extrinsic curvature, which leaves the background and quadratic
perturbation actions unchanged but contributes to  the source function of SIGWs.
For monochromatic and log-normal primordial curvature spectra,
we evaluate  finite-$\sigma$ effects without
the cubic interaction and study its contribution in the
large-$|\sigma|$ limit.
For the positive parameter values considered, finite-$\sigma$
effects suppress the spectrum over a broad range of scales relative to the result in the $\sigma\to\infty$ limit,
whereas the cubic interaction produces suppression at small
wavenumbers and enhancement at larger wavenumbers.

The paper is organized as follows.
In Sec.~\ref{Sec2}, we introduce the 2-DOF SCG model
and the minimally coupled matter field.
In Sec.~\ref{Sec3}, we derive the background
equations.
In Sec.~\ref{Sec4}, we derive the quadratic actions
for tensor and scalar perturbations and obtain the reduced scalar action.
In Sec.~\ref{Sec5}, we specify the radiation-dominated
background and study the evolution of the scalar mode
at finite
$\sigma$ and in the large-$|\sigma|$ limit.
In Sec.~\ref{Sec6},  we derive the equation of motion for SIGWs, construct the corresponding source function and kernel,  and evaluate the fractional energy density of SIGWs.
 We conclude in Sec.~\ref{Sec.7}.
Appendices \ref{app:QD}, \ref{app:general-degeneracy}, \ref{Cubic-action},  \ref{app:backreaction}  provide the general quadratic coefficients,
the proof of scalar kinetic degeneracy,
the scalar--scalar--tensor cubic action,
and estimates of non-linear backreaction.
Latin indices $i,j,k,\ldots$ are spatial indices.

\section{2-DOF spatially covariant gravity}\label{Sec2}

Spatially covariant gravity (SCG) provides a broad framework for modified gravity theories that are invariant under spatial diffeomorphisms rather than full spacetime diffeomorphisms. A class of SCG theories propagating at most three degrees of freedom (DOFs), namely, two tensor modes and one scalar mode, can be written as
\begin{equation}
    S=\int\mathrm{d}t\mathrm{d}^3xN\sqrt{h}\mathcal{L}(t,N,h_{ij},K_{ij},R_{ij},\nabla_i),
\end{equation}
where $N$ is the lapse function, $h_{ij}$ is the spatial metric, $R_{ij}$ is the spatial Ricci tensor and $\nabla_i$ is the covariant derivative compatible with the spatial metric. The extrinsic curvature is defined by
\begin{equation}
K_{ij}\coloneqq\frac{1}{2N}\left(\partial_t h_{ij}-\nabla_iN_j-\nabla_jN_i\right).
\end{equation} 
By imposing appropriate degeneracy conditions, the gravitational scalar mode can be eliminated, yielding subclasses that propagate only the two tensorial DOFs corresponding to the two polarization modes of gravitational waves. Through a gauge-recovering procedure, such 2-DOF SCG models can be recast as generally covariant scalar-tensor theories in which the timelike scalar field associated with the preferred foliation does not represent an independent propagating mode.

Recently, the perturbative construction of polynomial-type 2-DOF SCG theories was extended to cubic order around a cosmological background in Ref.~\cite{Yu:2026fgn}. Among the three resulting actions, the branch \(S^{\mathrm{A}}\) is singled out because it is compatible with the GR limit. We therefore take \(S^{\mathrm{A}}\) as the starting point of our analysis \footnote{We adopt units in which $8\pi G=1$ and normalize
the gravitational action so that its GR limit
takes the ADM form
$S_{\rm EH}=\frac12\int\mathrm dt\,\mathrm d^3x\,
N\sqrt h\,(K_{ij}K^{ij}-K^2+{}^{3}\!R)$,
up to boundary terms.
The overall factor of $1/2$ in $S^{\mathrm A}$
implements this convention.
All gravitational coefficient functions
throughout this work follow this normalization.}
\begin{equation}\label{SA}
    \begin{split}
S^{\mathrm{A}}&=\frac{1}{2}\int\mathrm{d}t\mathrm{d}^{3}xN\sqrt{h}\Big[\frac{C_{1}N}{C_{5}-2C_{2}N}\hat{K}^{ij}\hat{K}_{ij}+\frac{1}{3}\frac{C_{1}N}{1+C_{2}N}K^{2}+\Big(\frac{C_{3}}{N}+C_{4}\Big){}^{3}\!R+\Big(\frac{B_{1}}{N}+B_{2}\Big) \\
&+c_{1}^{(0;3,0)}\hat{K}_{ij}\hat{K}^{jk}\hat{K}_{k}^{i}+\frac{\big(D_{1}(C_{5}-2C_{2}N)+D_{2}\big)N^{2}}{(1+C_{2}N)(C_{5}-2C_{2}N)^{2}}\hat{K}_{ij}\hat{K}^{ij}K+\frac{1}{9}\frac{D_{1}N^{2}}{(1+C_{2}N)^{2}}K^{3} \\
&+\Big(\frac{6C_{4}D_{2}}{C_{1}C_{2}(C_{5}-2C_{2}N)}-2D_{3}\Big){}^{3}\!R^{ij}\hat{K}_{ij}+\frac{1}{3}D_{3}{}^{3}\!RK\Big],
 \end{split}
\end{equation}
where $B_n, C_n, D_n$ are arbitrary functions of time, independent of the lapse function, and satisfy
\begin{equation}
C_{4}\neq0,\qquad D_{1}D_{2}=0.
\end{equation}
The condition \(D_1D_2=0\) encodes the two branches \(D_1=0\) and \(D_2=0\). For convenience, we have decomposed the extrinsic curvature into its trace and traceless parts, 
\begin{equation}
    K_{ij}=\hat{K}_{ij}+\frac{1}{3}h_{ij}K,\qquad\hat{K}_{ij}h^{ij}\equiv0.
\end{equation}

We minimally couple the 2-DOF SCG model \(S^{\mathrm{A}}\) \footnote{Throughout this work, ``2-DOF'' refers to the perturbative DOF counting through cubic order around an FLRW background.} constructed in Ref.~\cite{Yu:2026fgn} to a \(k\)-essence matter field \(\chi\), whose perturbations source SIGWs.
 The total action is
\begin{equation}
    S=S^{\mathrm{A}}+S_{\mathrm{mat}},
\end{equation}
where the matter action is given by
\begin{equation}
    S_{\mathrm{mat}}=\int dt\,d^3x\,N\sqrt h\,P(\chi,X),
\end{equation}
with the standard kinetic term defined as
\begin{equation}
    X\equiv -\frac{1}{2}g^{\mu\nu}\partial_\mu\chi\partial_\nu\chi=
    \frac{1}{2N^2}\left(\dot\chi-N^i\partial_i\chi\right)^2-\frac12 h^{ij}\partial_i\chi\partial_j\chi.
\end{equation}
Here \(N^i\) is the shift vector. The scalar field can be divided into a homogeneous background and an inhomogeneous perturbation as
\begin{equation}
    \chi(t,\mathbf{x})=\bar{\chi}(t)+\delta\chi(t,\mathbf{x}).
\end{equation}

It is important to emphasise that although the pure gravitational sector \(S^{\mathrm{A}}\) has been constructed such that the gravitational scalar mode is absent through cubic order in perturbations, this property is not automatically preserved once matter is introduced. In general, matter coupling may modify the constraint structure and reactivate an additional gravitational scalar DOF. 

As will be shown below, the reduced quadratic scalar action contains only one propagating scalar DOF, which is associated with the \(k\)-essence field \(\chi\). Thus the theory with minimal coupling propagates one scalar mode and two tensor modes at the level of linear perturbations around an FLRW background.

\section{Background evolution}\label{Sec3}
To study cosmological perturbations around a spatially flat Friedmann--Lemaître--Robertson--Walker (FLRW) background, we parametrize the ADM variables as
\begin{equation}
N=\bar N e^{A},\quad\quad
N_i=\bar N a \partial_i B,\quad\quad
h_{ij}=a^2e^{2\zeta}\left(e^\gamma\right)_{ij},
\end{equation}
where the matrix exponential $e^{\gamma}$ is defined as
\begin{equation}
\left(e^\gamma\right)_{i j}=\delta_{i j}+\gamma_{i j}+\frac{1}{2} \gamma_i^k \gamma_{k j}+\frac{1}{6} \gamma_i^k \gamma_{k l} \gamma_j^l+\cdots.
\end{equation}
Here, $\bar{N}=\bar{N}(t)$ denotes the background value of the lapse function and $a=a(t)$ is the scale factor. The quantities $A, B, \zeta$ are scalar perturbations, while $\gamma_{i j}$ is the tensor perturbation. Vector perturbations are omitted since they do not contribute to the analysis considered below. Throughout the perturbative analysis, spatial indices are raised and lowered by the flat metric $\delta_{i j}$, e.g., $\gamma_i^j=\gamma_{i k} \delta^{k j}$. The tensor perturbation is symmetric, transverse, and traceless, i.e., $\partial^i \gamma_{i j}=0, ~\delta^{i j} \gamma_{i j}=0$. We also introduce the notation $\partial^i\equiv\delta^{i j} \partial_j, ~\partial^2\equiv\partial^i \partial_i$, for convenience. In the SCG formulation, the preferred foliation fixes the temporal slicing, which implies that we work in the so-called unitary gauge. Unlike in GR, the temporal diffeomorphism freedom is not available to impose an alternative gauge condition such as the Newtonian gauge. Here the unitary gauge refers to the preferred foliation of the gravitational sector and does not require $\delta\chi=0$, since $\chi$ is an independent matter scalar field.

The action expanded to first order in perturbations takes the form
\begin{equation}
  \begin{split}
        S^{(1)}=\int \mathrm d \eta\mathrm d^3x ~a^4\left(\mathcal{E}_A A+\mathcal{E}_\zeta \zeta +\mathcal{E}_{\chi}\delta \chi\right),
  \end{split}
\end{equation}
where
\begin{equation}
    \begin{split}
         \mathcal{E}_A&=\bar{P}-\frac{\bar{\chi}'^2\bar{P}_X}{a^2}+\frac12B_2-\frac32\left(\frac{\mathcal{H}}{a}\right)^2\frac{C_1C_2\bar{N}^2}{(1+C_2\bar{N})^2}   
        -\left(\frac{\mathcal{H}}{a}\right)^3\frac{3C_2D_1\bar{N}^3}{(1+C_2\bar{N})^3},\\
\mathcal{E}_{\zeta}&=3\bar{P}+\frac32\left(\frac{B_1}{\bar{N}}+B_2\right)
+\frac92\left(\frac{\mathcal{H}}{a}\right)^2\frac{C_1\bar{N}}{(1+C_2\bar{N})
}+\frac92\left(\frac{\mathcal{H}}{a}\right)^3\frac{D_1\bar{N}^2}{(1+C_2\bar{N})^2}
       \\
        &~~~
        -\frac{1}{a^4}\partial_\eta\left[\left(\frac{\mathcal{H}}{a}\right)\frac{3a^3C_1\bar{N}}{(1+C_2\bar{N})} 
      +\frac92\left(\frac{\mathcal{H}}{a}\right)^2\frac{a^3D_1\bar{N}^2}{(1+C_2\bar{N})^2} \right],\\
\mathcal{E}_{\chi}&=\bar{P}_{\chi}
-\frac{1}{a^4}\partial_\eta\left(a^2\bar{\chi}'\bar{P}_X  \right).
    \end{split}
\end{equation}
Here and below, all coefficient functions are understood to be
evaluated on the homogeneous background. We have introduced the
conformal time $\eta$ through  $a\mathrm d\eta=\bar{N}\mathrm d t$ and the conformal Hubble parameter $\mathcal H\equiv\frac{a'}{a}$. A prime denotes differentiation with respect to conformal time $\eta$.

Requiring the first-order action to vanish gives the background equations
\begin{equation}\label{L1-eom}
    \begin{split}
        3\mathcal{H}^2=a^2\bar{\rho}_{\text{eff}}
        ,\quad\quad
        \mathcal H^2+2\mathcal H'=-a^2\bar{p}_{\text{eff}},
    \end{split}
\end{equation}
where the effective energy density and pressure are defined as
\begin{equation}
\begin{split}\label{bar-p-rho}
\bar{\rho}_{\text{eff}}&=\bar\rho_k+3\left(\frac{\mathcal{H}}{a}\right)^2\left(1+\frac{C_1C_2\bar{N}^2}{2(1+C_2\bar{N})^2}\right)-\frac{B_2}{2}+
\left(\frac{\mathcal{H}}{a}\right)^3\frac{3C_2D_1\bar{N}^3}{(1+C_2\bar{N})^3},\\
\bar{p}_{\text{eff}}&=\bar{p}_k+3\left(\frac{\mathcal{H}}{a}\right)^2\left(1+\frac{C_1\bar{N}}{2(1+C_2\bar{N})}\right) 
    +\frac12\left(\frac{B_1}{\bar{N}}+B_2\right)
    +
\frac32\left(\frac{\mathcal{H}}{a}\right)^3\frac{D_1\bar{N}^2}{(1+C_2\bar{N})^2}
    \\
    &~~~
    -\frac{2}{a^4}\partial_\eta\left[a^3\left(\frac{\mathcal{H}}{a}\right)\left(1+\frac{C_1\bar{N}}{2(1+C_2\bar{N})}\right) +a^3\left(\frac{\mathcal{H}}{a}\right)^2\frac{3D_1\bar{N}^2}{4(1+C_2\bar{N})^2} \right].
\end{split}
\end{equation}
The background energy density and pressure of the $k$-essence
sector are
\begin{equation} 
\bar\rho_k=\frac{\bar{\chi}'^2\bar{P}_X}{a^2}-\bar{P}=2\bar{X}\bar{P}_{X}-\bar{P},
        \qquad\qquad
\bar{p}_k=\bar{P},
\end{equation}
where the background kinetic term is
\begin{equation}
\bar X=\frac{\bar\chi'^2}{2a^2}.
\end{equation}

The two effective Friedmann equations imply the continuity equation
\begin{equation}
\bar\rho_{\rm eff}'
+
3\mathcal H
\left(
\bar\rho_{\rm eff}+\bar p_{\rm eff}
\right)
=0.
\end{equation}
We define the effective equation-of-state parameter as
\begin{equation}
w_{\rm eff}\equiv
\frac{\bar p_{\rm eff}}{\bar \rho_{\rm eff}}.
\end{equation}
If $w_{\rm eff}$ is constant, the effective energy density evolves according to
\begin{equation}
\bar\rho_{\rm eff}
\propto
a^{-3(1+w_{\rm eff})}.
\end{equation}
Combining this relation with the first Friedmann equation gives
\begin{equation}
a(\eta)
\propto
\eta^{\frac{2}{1+3w_{\rm eff}}},
\qquad
\mathcal H=
\frac{2}
{(1+3w_{\rm eff})\eta},
\end{equation}
for $w_{\rm eff}\neq-1/3$. It follows that
\begin{equation}
\mathcal H'=
-\frac{1+3w_{\rm eff}}{2}
\mathcal H^2.
\end{equation}
For the \(k\)-essence matter field, the background equation-of-state parameter
is
\begin{equation}\label{wk}
    w_k\equiv \frac{\bar p_k}{\bar \rho_k}
    =
    \frac{\bar P}{2\bar X \bar{P}_X-\bar{P}}.
\end{equation}
Here, \(w_k\) describes the matter field alone,
whereas \(w_{\rm eff}\) characterizes the total effective background
evolution, accounting for both the matter field and
modifications to gravity. The squared intrinsic sound speed of the $k$-essence matter field is
\begin{equation}
    c_s^2=
    \frac{\bar{P}_X}{\bar{P}_X+2\bar{X} \bar{P}_{XX}}.
\end{equation}

In this work, we focus on 
\begin{equation}
    w_{\rm eff}=\frac13,
\end{equation}
so that the total effective background expansion coincides with that
of a standard radiation-dominated universe. The effective radiation-dominated expansion history is
characterized by
\begin{equation}
a(\eta)\propto\eta,
\qquad
\mathcal H=\frac1\eta,
\qquad
\mathcal H'=-\mathcal H^2.
\end{equation}
For simplicity, we
further assume that the \(k\)-essence matter field itself also
behaves as radiation at the background level, $ w_k=\bar p_k/\bar\rho_k=1/3$. This condition implies $2\bar P=\bar X\bar P_X $.
A simple choice that satisfies this relation identically is
\begin{equation}\label{P}
    P(\chi,X)=\lambda_0X^2,
\end{equation}
where \(\lambda_0\) is a constant. For this choice, one finds
\begin{equation}
\bar{\rho}_k=3\lambda_0\bar{X}^2,
\qquad
\bar{p}_k=\lambda_0\bar{X}^2,
\qquad
c_s^2=\frac{1}{3}.
\end{equation}
Therefore, the \(k\)-essence scalar field has the equation of state and intrinsic
sound speed of radiation. The evolution of the coupled scalar
mode is determined by the reduced equations derived below.

\section{Quadratic actions}\label{Sec4}
In this section, we derive the quadratic actions for tensor and scalar perturbations. We first derive the quadratic tensor action for the general \(S^{\mathrm A}\) theory and then select a subclass whose quadratic tensor action coincides with that of GR. Next, we analyze the scalar constraints in Fourier space, establish the degeneracy of the reduced kinetic matrix, and derive the equation of motion for the propagating scalar combination.

\subsection{Tensor perturbations}
The quadratic action for the tensor perturbations $\gamma_{i j}$ takes the form
\begin{equation}
    S^{(2)}_\gamma=\int \mathrm{d} \eta 
    \mathrm{d}^3 x
    \left(\mathcal{G}_\gamma\left(\gamma_{i j}^{\prime}\right)^2 -\mathcal{W}_{\gamma}  (\partial_k\gamma_{i j})^2\right),
\end{equation}
where the coefficients are given by
\begin{equation}
    \begin{split}
       \mathcal{G}_\gamma&=\frac{a^2}{8}\Bigg[\left(\frac{C_1\bar{N}}{C_5-2C_2\bar{N}}\right)
       +\frac{3\mathcal{H}}{a}\frac{\left(D_2+(C_5-2C_2\bar{N})D_1\right)\bar{N}^2}{(C_5-2C_2\bar{N})^2(1+C_2\bar{N})}\Bigg],\\[3pt]       
       \mathcal{W}_\gamma=& \frac{a^2}{8}\Bigg[\left(\frac{C_3}{\bar{N}}+C_4+\frac{\mathcal H D_3}{a}\right)
        -\frac{1}{a^2}\partial_{\eta}\left(aD_3-\frac{3aC_4D_2}{C_1C_2(C_5-2C_2\bar{N})}\right)\Bigg].
    \end{split}
\end{equation}
The tensor propagation speed is defined by
\begin{equation}
c_{T}^{2}\equiv\frac{\mathcal{W}_{\gamma}}
{\mathcal{G}_{\gamma}}.
\end{equation}

As a toy model, we consider a subclass of the action \(S^A\) defined by the following choice of parameters
\begin{equation}\label{choice-CD}
C_{1}=-2C_{2}\neq0,
\qquad
C_{3}=C_{5}=D_{1}=D_{2}=D_{3}=0,
\qquad
C_{4}=1.
\end{equation}
The gravitational action becomes
\begin{equation} 
\begin{split} \label{toy-model}
S_{\mathrm{gra}}& = \frac12 \int \mathrm{d}t\,\mathrm{d}^{3}x\, N\sqrt{h}\, \bigg[  
K^{ij}K_{ij}-K^{2}+{}^{3}\!R+\frac{2}{3(1+C_{2}N)}K^{2} \\
&\ \ +\left(\frac{B_{1}}{N}+B_{2}\right) +c_{1}^{(0;3,0)} \hat K_{i}^{j}\hat K_{j}^{k}\hat K_{k}^{i} \bigg].
\end{split} 
\end{equation}
For this parameter choice, only \(\hat K_{ij}\hat K^{ij}\) and \({}^{3}R\) contribute to the quadratic tensor action, with the same coefficients as in GR. Hence, on the same FLRW background, the tensor kinetic normalization and propagation speed agree exactly with those of GR:
\begin{equation} \mathcal{G}_\gamma = \mathcal{W}_\gamma = \frac{a^2}{8}, \qquad c_T^2=1. 
\end{equation} 

The first three terms in Eq.~\eqref{toy-model} reproduce the ADM form of the Einstein--Hilbert Lagrangian up to boundary terms. The additional \(K^2\) term does not affect the quadratic tensor action and vanishes as $ C_2N \to\infty$. The operator proportional to \(c_{1}^{(0;3,0)}\) likewise does not contribute at quadratic order, but remains a characteristic cubic interaction of our model. After restoring spacetime diffeomorphism invariance, the terms involving \(B_1\) and \(B_2\) correspond to a cuscuton kinetic term and a scalar-field potential, respectively.

\subsection{Scalar perturbations}

 To examine how the matter coupling affects the scalar constraint structure, we first retain the general coefficients of \(S^{\mathrm A}\). We adopt the parameter choice in Eq.~\eqref{choice-CD}
in Sec.~\ref{Sec5}, where we study the evolution of
scalar perturbations during radiation domination.

In Fourier space, the quadratic action for the scalar perturbations can be written as 
\begin{equation}\label{QC1}
\begin{split}
S_{ss}^{(2)}
&=
\int d\eta \frac{d^3k}{(2\pi)^3}a^4
\Big(
\zeta_k' Q_{\zeta'\zeta'}\zeta_{-k}'
+\zeta_k' Q_{\zeta'\zeta}\zeta_{-k}
+\zeta_k Q_{\zeta\zeta}\zeta_{-k}
+A_k Q_{AA}A_{-k}
+A_k Q_{A\zeta}\zeta_{-k}
\\
&
+A_k Q_{A\zeta'}\zeta_{-k}'
+A_k Q_{AB}B_{-k}
+B_k Q_{B\zeta}\zeta_{-k}
+B_k Q_{B\zeta'}\zeta_{-k}'
+B_k Q_{BB}B_{-k}
\\
&
+\delta\chi'_k Q_{\chi'\chi'}\delta\chi'_{-k}
+\delta\chi'_k Q_{\chi'\chi}\delta\chi_{-k}
+\delta\chi_k Q_{\chi\chi}\delta\chi_{-k}
+\delta\chi'_k Q_{\chi'\zeta}\zeta_{-k}
+\delta\chi_k Q_{\chi\zeta}\zeta_{-k}
\\
&
+\delta\chi'_k Q_{\chi' A}A_{-k}
+\delta\chi_k Q_{\chi A}A_{-k}
+\delta\chi_k Q_{\chi B}B_{-k}
\Big).
\end{split}
\end{equation}
The explicit expressions for the coefficients \(Q_{nn}\) are collected in Appendix~\ref{app:QD}. As shown in Appendix~\ref{app:QD}, these terms generally involve the coefficients \(C_{n}\) and \(D_{n}\), while \(B_1\) and \(B_2\) have already been eliminated by using the background equations. The coefficient \(c_1^{(0;3,0)}\) does not enter at this order, but appears in the cubic action, as discussed in Sec.~\ref{Sec6}.

The scalar perturbations in the lapse function and shift vector, i.e., \(A\) and \(B\), are auxiliary variables whose equations of motion provide two constraint equations. 
Assuming  the lapse--shift constraint matrix is invertible, namely, \(\Delta\equiv4Q_{AA}Q_{BB}-Q_{AB}^2\neq0\), the solutions of the constraint
equations are
\begin{equation}
\begin{aligned}
A_k & =\frac{-2 Q_{B B}\left(Q_{A \zeta^{\prime}} \zeta_k^{\prime}+Q_{A \zeta} \zeta_k + Q_{\chi' A}\delta \chi'_k+ Q_{\chi A}\delta \chi_k\right)
+Q_{A B}\left(Q_{B \zeta^{\prime}} \zeta_{k}^{\prime}+Q_{B \zeta} \zeta_k+Q_{\chi B}\delta \chi_k\right)}{4Q_{AA}Q_{BB}-Q_{AB}^2}, \\[4pt]
B_k & =\frac{-2 Q_{A A}\left(Q_{B \zeta^{\prime}} \zeta_{k}^{\prime}+Q_{B \zeta} \zeta_k+Q_{\chi B}\delta \chi_k\right)
+Q_{A B}\left(Q_{A \zeta^{\prime}} \zeta_k^{\prime}+Q_{A \zeta} \zeta_k + Q_{\chi' A}\delta \chi'_k+ Q_{\chi A}\delta \chi_k\right)}{4Q_{AA}Q_{BB}-Q_{AB}^2}.
\end{aligned}
\end{equation}
Substituting these solutions back into the quadratic action gives the
reduced action for the coupled scalar perturbations \(\zeta\) and
\(\delta\chi\),
\begin{equation}\label{QC2}
\begin{split}
    S_{ss}^{(2)}
=&
\int d\eta \frac{d^3 k}{(2\pi)^3}
a^4\Big(
\zeta_k'D_{\zeta'\zeta'}\zeta_{-k}'
+\zeta_kD_{\zeta'\zeta}\zeta_{-k}'
+\zeta_kD_{\zeta\zeta}\zeta_{-k}\\
&+\delta\chi_k'D_{\chi'\chi'}\delta\chi_{-k}'
+\delta\chi_kD_{\chi'\chi}\delta\chi_{-k}'
+\delta\chi_kD_{\chi\chi}\delta\chi_{-k}
\\
&
+\zeta_k'D_{\zeta'\chi'}\delta\chi_{-k}'
+\zeta_k D_{\zeta\chi'}\delta\chi_{-k}'
+\zeta'_kD_{\zeta'\chi}\delta\chi_{-k}
+\zeta_kD_{\zeta\chi}\delta\chi_{-k}
\Big).
\end{split}
\end{equation}
The explicit expressions for the reduced coefficients \(D_{nn}\) are
given in Appendix~\ref{app:QD}.

As shown in Appendix~\ref{app:general-degeneracy},
for \(\Delta\neq0\), 
the kinetic matrix of the coupled scalar system is degenerate
\begin{equation}\label{detK}
    \det\mathbf K = 4D_{\zeta'\zeta'}D_{\chi'\chi'} -D_{\zeta'\chi'}^2 = 0,
\end{equation}
for arbitrary \(B_n\), \(C_n\), and \(D_n\) within
\(S^{\mathrm{A}}\), and for a general minimally coupled
\(k\)-essence Lagrangian \(P(\chi,X)\). 
Consequently, although the scalar sector is
described by the two variables \(\zeta_k\) and \(\delta\chi_k\),  its  kinetic terms can be expressed in terms of  a single scalar combination, provided that \(D_{\chi'\chi'}\neq0\). A convenient choice for the dynamical combination is
\begin{equation}
\varphi_k\equiv\delta\chi_k-r\zeta_k,
\qquad
r\equiv-\frac{D_{\zeta'\chi'}}{2D_{\chi'\chi'}} .
\label{general-Phi}
\end{equation}
Here,
\(\varphi_k\) represents the propagating scalar degree of freedom. In general, $r$ may depend on time. 
Substituting this field redefinition into the reduced action
and integrating the terms containing $\zeta_k'$ by parts gives
\begin{equation}\label{QC3}
\begin{split}
S_{ss}^{(2)}={}
&
\int d\eta\frac{d^3k}{(2\pi)^3}a^4
\Big(
\mathcal D_{\varphi'\varphi'}\varphi_k'\varphi_{-k}'
+\mathcal D_{\varphi\varphi'}\varphi_k\varphi_{-k}'
+\mathcal D_{\varphi\varphi}\varphi_k\varphi_{-k}
\\
&
+\lambda_1\zeta_k\varphi_{-k}'
+\lambda_2\zeta_k\varphi_{-k}
+\lambda_3\zeta_k\zeta_{-k}
\Big),
\end{split}
\end{equation}
where  \(\mathcal D_{nn}\) are
shown in Appendix~\ref{app:QD}, and
\begin{equation}\label{lambda123}
\lambda_1\equiv
\mathcal D_{\zeta\varphi'}-\mathcal D_{\zeta'\varphi},
\qquad
\lambda_2\equiv
\mathcal D_{\zeta\varphi}
-\left(\mathcal D_{\zeta'\varphi}\right)'
-4\mathcal H\mathcal D_{\zeta'\varphi},
\qquad
\lambda_3\equiv
\mathcal D_{\zeta\zeta}
-\frac12\left(\mathcal D_{\zeta'\zeta}\right)'
-2\mathcal H\mathcal D_{\zeta'\zeta}.
\end{equation}
Thus, $\zeta$ carries no independent kinetic term.
Provided that $\lambda_3\neq0$, the equation of motion with respect to $\zeta_k$ is an algebraic constraint
\begin{equation}\label{eq:ABC-B2-zero}
2\lambda_3\zeta_k+\lambda_1\varphi_k'+\lambda_2\varphi_k=0,
\end{equation}
which
gives
\begin{equation}
\zeta_k=-\frac{\lambda_1\varphi_k'+\lambda_2\varphi_k}{2\lambda_3}.
\end{equation}
Substituting the algebraic solution for
$\zeta_k$ back into the action \eqref{QC3} leaves an effective action involving only one propagating scalar mode
\begin{equation}\label{QC4}
S_{\varphi}^{(2)}
=
\int d\eta\frac{d^3k}{(2\pi)^3}a^4
\left[
\widehat{\mathcal D}_{\varphi'\varphi'}\varphi_k'\varphi_{-k}'
+\widehat{\mathcal D}_{\varphi\varphi'}\varphi_k\varphi_{-k}'
+\widehat{\mathcal D}_{\varphi\varphi}\varphi_k\varphi_{-k}
\right],
\end{equation}
where
\begin{equation}
\widehat{\mathcal D}_{\varphi'\varphi'}
=\mathcal D_{\varphi'\varphi'}-\frac{\lambda_1^2}{4\lambda_3},
\qquad
\widehat{\mathcal D}_{\varphi\varphi'}
=\mathcal D_{\varphi\varphi'}-\frac{\lambda_1\lambda_2}{2\lambda_3},
\qquad
\widehat{\mathcal D}_{\varphi\varphi}
=\mathcal D_{\varphi\varphi}-\frac{\lambda_2^2}{4\lambda_3}.
\end{equation}
We restrict attention to branches on which
$\widehat{\mathcal D}_{\varphi'\varphi'}\neq0$,
so that $\varphi_k$ retains a nonvanishing quadratic kinetic term.

\section{Evolution of the scalar mode during radiation domination}\label{Sec5}
We specialize to a radiation-dominated era with
$P=\lambda_0X^2$ and define
\begin{equation}
Y\equiv\lambda_0a^2\bar X^2.
\end{equation}
The background matter equation gives
\begin{equation}\label{eq:matter}
\left(a^2\bar P_X\bar\chi'\right)'
=
\left(\lambda_0\bar\chi'^3\right)'=0,
\qquad
\bar\chi'=\mathrm{const}.
\end{equation}
It follows that
\begin{equation}\label{YH1}
\frac{\bar X'}{\bar X}=-2\mathcal H,
\qquad
Y'=-2\mathcal H Y.
\end{equation}
For an effective radiation-dominated era,
$w_{\rm eff}=1/3$, one further has $\mathcal H'=-\mathcal H^2$.
Together with
Eq.~\eqref{YH1}, this implies
\begin{equation}
\left(\frac{Y}{\mathcal H^2}\right)'=0.
\label{eq:Y-H-ratio}
\end{equation}
We shall also use the identities $\bar\chi'^2=2a^2\bar X$ and $
\lambda_0\bar\chi'^2\bar X=2Y$, which follow directly from the definitions of $\bar X$ and $Y$.

We consider the parameter choice \eqref{choice-CD}
while retaining the background functions $B_1$ and $B_2$.
Introducing
\begin{equation}
\sigma\equiv C_2\bar N,
\end{equation}
we write the background equations \eqref{L1-eom} as
\begin{equation}
Y
=
\mathcal H^2
\left(\frac{\sigma}{1+\sigma}\right)^2
+\frac{a^2B_2}{6},
\label{eq:bg1}
\end{equation}
and
\begin{equation}
Y+\frac{\sigma}{1+\sigma}
\left(\mathcal H^2+2\mathcal H'\right)
+\frac{2\mathcal H\sigma'}{(1+\sigma)^2}
+\frac{a^2}{2}
\left(\frac{B_1}{\bar N}+B_2\right)
=0.
\label{eq:bg2}
\end{equation}
For the radiation-dominated era considered here,
Eqs.~\eqref{eq:bg1} and~\eqref{eq:bg2} determine
$B_2$ and $B_1/\bar N$ once $\sigma(\eta)$ and the
constant value of $Y/\mathcal H^2$ are specified. 

In the following, we first derive the scalar constraints
and evolution equation for general $\sigma(\eta)$, with $B_1$ and $B_2$ satisfying the
background equations but without
imposing $B_2=0$.
We then specialize to $B_2=0$ for the
large-$|\sigma|$ analysis and for the transfer functions
and numerical calculations in Sec.~\ref{Sec6}. In this case, Eq.~\eqref{eq:bg1} gives
 \begin{equation}
Y=\mathcal H^2
\left(\frac{\sigma}{1+\sigma}\right)^2.
\label{eq:YB-B2-zero}
\end{equation}
Since $Y/\mathcal H^2$ is constant during radiation domination,
this relation implies $\sigma'=0$ for finite $\sigma\neq 0,-1$. The remaining background equation~\eqref{eq:bg2}
then fixes $B_1/\bar N$. Thus, for the $B_2=0$ backgrounds considered below, $\sigma$ is constant and $Y$ is related to $\mathcal H$ by Eq.~\eqref{eq:YB-B2-zero}.

\subsection{Scalar dynamics at finite $\sigma$}

The background functions $B_1$ and $B_2$ have already
been eliminated from the quadratic coefficients
using the background equations.
The coefficients entering the algebraic curvature constraint are
\begin{equation}
    \begin{split}
\lambda_1
=
\frac{4Y}{a^2\bar\chi'\Xi}\mathcal C_{\sigma} ,
\qquad\qquad
\lambda_2
=
-\mathcal R_{\sigma}\lambda_1,
\qquad\qquad
\lambda_3
=
\frac{\mathcal L_{\sigma}}
{3\mathcal H^2a^2\sigma^3\Xi},
    \end{split}
\end{equation}
where
\begin{equation}
\begin{split}
\Xi
&\equiv
\mathcal H^2\sigma^3-2Y(1+\sigma)^2,
\\
\mathcal R_\sigma
&\equiv
\frac{2Y(1+\sigma)}{\mathcal H\sigma},
\\
\mathcal L_\sigma
&\equiv
\Xi\mathcal P_\sigma+2Y\mathcal C_\sigma^2,
\end{split}
\end{equation}
and 
\begin{equation}
\begin{split}
\mathcal C_\sigma
&=
3\mathcal H^2\sigma^2(2\sigma+1)
-6Y(1+\sigma)^3
-k^2(1+\sigma)^2-3\mathcal H \sigma \sigma',
\\[3pt]
\mathcal P_{\sigma}
&=
k^4(1+\sigma)^2
+6Yk^2(1+\sigma)^2(\sigma+2)
+36Y^2(1+\sigma)^3
\\
&-
3\mathcal H^2k^2\sigma^2(2\sigma+3)
-18\mathcal H^2Y\sigma^2(2\sigma+1)
+3\mathcal H \sigma(k^2+6Y)\sigma'.
\end{split}
\end{equation}
Throughout, we restrict attention to nonzero Fourier modes
and to branches on which the constraint reduction is non-singular.
In particular, the lapse--shift constraints require $\Xi\neq0$.

Provided that $\lambda_3\neq0$, the nondynamical curvature
perturbation is determined algebraically as
\begin{equation}
\zeta_k
=
-\frac{
6Y\mathcal H^2\sigma^3\mathcal C_{\sigma}
}{
\bar\chi'\mathcal L_{\sigma}
}
\left(
\varphi_k'-\mathcal R_{\sigma}\varphi_k
\right).
\end{equation} 
The matter, lapse, and shift perturbations can then
be expressed as
\begin{equation}
\delta\chi_k
=
\varphi_k
+
\frac{\bar\chi'(1+\sigma)}{\mathcal H \sigma}\zeta_k,
\end{equation}
\begin{equation}\label{eq:A-finite}
\begin{split}
A_k={}&
\frac{1+\sigma}{\mathcal H \sigma}\zeta_k'
+\frac{2\mathcal H Y\sigma^2(1+\sigma)}{\bar\chi'\Xi}\varphi_k
-\frac{2Y(1+\sigma)^2}{\bar\chi'\Xi}\varphi_k'
\\
&+\frac{(1+\sigma)^2}{3\sigma\Xi}
\left[
k^2\sigma+6Y(\sigma^2-\sigma-1)+\frac{6Y\sigma'}{\mathcal H\sigma}
\right]\zeta_k,
\end{split}
\end{equation}
and
\begin{equation}
\begin{split}\label{eq:B-finite}
B_k={}&
\frac{6Y[\mathcal H^2\sigma^3-2Y(1+\sigma)^3]}
{\bar\chi'k^2\Xi}\varphi_k
+\frac{6\mathcal H Y\sigma^2(1+\sigma)}
{\bar\chi'k^2\Xi}\varphi_k'
\\
&-
\frac{
(1+\sigma)[\mathcal H^2\sigma^2(k^2\sigma-12Y\sigma-6Y)
+12Y^2(1+\sigma)^3+6\mathcal HY \sigma\sigma' ]
}{k^2\mathcal H \sigma\Xi}
\zeta_k ,
\end{split}
\end{equation}
which follow directly from
the algebraic lapse and shift constraints and the field redefinition.

Varying the reduced scalar action~\eqref{QC4} yields
\begin{equation}
\varphi_k''
+\Gamma_k\varphi_k'
+\Omega_k^2\varphi_k=0,
\label{eq:Phi-eom}
\end{equation}
where
\begin{equation}
\begin{split}
\Gamma_k
&=
-2\mathcal H+\frac{3\sigma'}{\sigma}
+\frac{\mathcal P_\sigma'}{\mathcal P_{\sigma}}
-\frac{\mathcal L_\sigma'}{\mathcal L_{\sigma}},
\\
\Omega_k^2
&=-
\mathcal R_{\sigma}\Gamma_k
-\mathcal R_{\sigma}'
-\mathcal R_{\sigma}^2
+\frac{(k^2+6Y)\mathcal L_{\sigma}}
{3\mathcal H^2\sigma^3\mathcal P_{\sigma}}.
\end{split}
\end{equation}

\subsection{Large-$\lvert\sigma\rvert$ limit}

We now impose the background relation~\eqref{eq:YB-B2-zero}
with $\sigma'=0$.
Having solved the constraints at finite $\sigma$, we
expand the solutions for $\lvert\sigma\rvert\gg1$ at fixed
$k$ and conformal time.
The background relation~\eqref{eq:YB-B2-zero} gives
\begin{equation}
Y
=
\mathcal H^2
\left[
1-\frac{2}\sigma 
+\mathcal O\!\left(\sigma^{-2}\right)
\right].
\end{equation}
The $\mathcal O(\sigma^{-1})$ correction to $Y/\mathcal H^2$
contributes to the leading nonvanishing
terms of the constraint coefficients $\lambda_{1,2,3}$
and must therefore
be retained. Expanding these coefficients gives
\begin{equation}
\lambda_1=-\frac{4(k^2+3\mathcal H^2)}{a^2\bar\chi' \sigma}
+\mathcal O(\sigma^{-2}),
\qquad
\lambda_2=\frac{8\mathcal H(k^2+3\mathcal H^2)}{a^2\bar\chi' \sigma}
+\mathcal O(\sigma^{-2}),
\qquad
\lambda_3=\frac{\mathcal Q_k}{3a^2\mathcal H^2\sigma}
+\mathcal O(\sigma^{-2}),
\end{equation}
with $\mathcal Q_k=k^4+3 k^2 \mathcal H^2+18 \mathcal H^4$. Although all three coefficients $\lambda_{1,2,3}$  vanish in the limit, the algebraic
solution for $\zeta_k$ remains finite
\begin{equation}\label{eq:zeta-large-sigma}
    \zeta_k
=
\frac{
6\mathcal H^2(k^2+3\mathcal H^2)
}{
\bar\chi'\mathcal Q_k
}
\left(
\varphi_k'-2\mathcal H\varphi_k
\right)+\mathcal O\left(\sigma^{-1}\right).
\end{equation}
The remaining constraint solutions reduce to
\begin{equation}
\begin{split}
&\delta\chi_k
=
\varphi_k
+
\frac{\bar\chi'}{\mathcal H}\zeta_k
+\mathcal O\left(\sigma^{-1}\right),
\\
&A_k={}\frac{\zeta_k'}{\mathcal H}
+\frac{2\mathcal H}{\bar\chi'}\varphi_k+2\zeta_k+\mathcal O\left(\sigma^{-1}\right),
\\
&
B_k
=\frac{6\mathcal H}{\bar\chi'k^2}\varphi_k'
-\frac{6\mathcal H^2}{\bar\chi'k^2}\varphi_k
-\frac{\zeta_k}{\mathcal H}+\mathcal O\left(\sigma^{-1}\right)
.
\end{split}
\end{equation}

At the same order, the coefficients of the scalar evolution
equation satisfy
\begin{equation}
   \begin{split}
     \Gamma_k
    =
    \mathcal O\left(\sigma^{-1}\right),
  \qquad\qquad
    \Omega^2_k= \frac{k^2}{3}
    +\mathcal O\left(\sigma^{-1}\right).
   \end{split}
\end{equation}
The leading-order equation is therefore
\begin{equation}
\varphi_k''+\frac{k^2}{3}\varphi_k\simeq0.
\label{eq:eom-varphi-large-u}
\end{equation}
The general solution of this equation is
\begin{equation}
\varphi_k(\eta)
=
\alpha_k\cos\left(\frac{k\eta}{\sqrt3}\right)
+
\beta_k\sin\left(\frac{k\eta}{\sqrt3}\right).
\end{equation}
Here the origin of conformal time is chosen such that
$\mathcal H=1/\eta$.
By Eq.~\eqref{eq:zeta-large-sigma}, the cosine mode produces
a curvature perturbation proportional to $\eta^{-1}$ in the
superhorizon limit and is excluded by regularity.
The regular solution is thus
\begin{equation}
\varphi_k(\eta)
=
\beta_k\sin\left(\frac{k\eta}{\sqrt3}\right),
\label{eq:varphi-sine-solution}
\end{equation}
whose amplitude is fixed by the primordial curvature normalization,
which we implement below using scalar transfer functions.

\section{Scalar-induced gravitational waves}\label{Sec6}

We decompose the tensor perturbation into its two polarization
modes,
\begin{equation}
\gamma_{ij}(\eta,\boldsymbol{x})
=
\sum_{\lambda=+,\times}
\int\frac{d^3k}{(2\pi)^3}\,
e^\lambda_{ij}(\hat{\boldsymbol{k}})
\gamma_{\boldsymbol{k},\lambda}(\eta)
e^{i\boldsymbol{k}\cdot\boldsymbol{x}},
\label{eq:tensor-polarization-decomposition}
\end{equation}
where the transverse and traceless polarization tensors satisfy $e^\lambda_{ij}(\hat{\boldsymbol{k}})
e^{\lambda'ij}(-\hat{\boldsymbol{k}})
=
\delta_{\lambda\lambda'}$.

Varying the quadratic tensor action together with the cubic
scalar--scalar--tensor action gives
\begin{equation}
\begin{split}
   & \gamma_{\boldsymbol{k},\lambda}^{\prime \prime}+\frac{\mathcal{G}_\gamma^{\prime}}{\mathcal{G}_\gamma} \gamma_{\boldsymbol{k},\lambda}^{\prime}+c_T^2 k^2 \gamma_{\boldsymbol{k},\lambda}=
    4\mathcal S_{\boldsymbol{k},\lambda} ,
   \end{split}
\end{equation}
where the source term is given by
\begin{equation}
    \begin{split}
 \mathcal S_{\boldsymbol{k},\lambda}= 
 \int \frac{d^3 q}{(2 \pi)^3} e_\lambda^{i j}(\hat{\mathbf{k}}) q_i q_j  &\frac{1}{8 \mathcal{G}_\gamma} \Bigg[
\Big(
\mathcal F_{BB}  B_{\boldsymbol{q}}B_{\boldsymbol{k-q}}
+\mathcal F_{B\zeta} 
B_{\boldsymbol{q}}\zeta_{\boldsymbol{k-q}}
+\mathcal F_{B\zeta'}B_{\boldsymbol{q}} \zeta'_{\boldsymbol{k-q}}
\\
&
+\mathcal F_{AB}  A_{\boldsymbol{q}}B_{\boldsymbol{k-q}}
+\mathcal F_{A\zeta} A_{\boldsymbol{q}} \zeta_{\boldsymbol{k-q}}
+\mathcal F_{\zeta\zeta}\zeta_{\boldsymbol{q}}\zeta_{\boldsymbol{k-q}}
+\mathcal F_{\zeta'\zeta}\zeta'_{\boldsymbol{q}}\zeta_{\boldsymbol{k-q}}
\\
&
+\mathcal F_{B\chi} B_{\boldsymbol{q}}\delta\chi_{\boldsymbol{k-q}} 
+     \mathcal F_{\chi\chi} \delta\chi_{\boldsymbol{q}} \delta\chi_{\boldsymbol{k-q}} \Big)\\
&- \partial_{\eta}
\Big(
\mathcal G_{BB} B_{\boldsymbol{q}}B_{\boldsymbol{k-q}}
+\mathcal G_{B\zeta} B_{\boldsymbol{q}}\zeta_{\boldsymbol{k-q}}
+\mathcal G_{B\zeta'} B_{\boldsymbol{q}}\zeta'_{\boldsymbol{k-q}}\\
&
+\mathcal G_{AB} A_{\boldsymbol{q}}B_{\boldsymbol{k-q}}
+\mathcal G_{A\zeta} A_{\boldsymbol{q}}\zeta_{\boldsymbol{k-q}}
+\mathcal G_{\zeta\zeta}\zeta_{\boldsymbol{q}}\zeta_{\boldsymbol{k-q}} 
\Big)\Bigg].
\end{split}
\end{equation}
The coefficients $\mathcal F_{nn}$ and $\mathcal G_{nn}$ are given in Appendix~\ref{Cubic-action}. 

 We solve for the tensor perturbation $\gamma_{\mathbf{k}}(\eta)$ by using the Green's function method. For a general time-dependent coefficient \(\mathcal G_\gamma\), 
 the solution can be written as
\begin{equation}
\sqrt{\mathcal G_\gamma(\eta)}\,
\gamma_{\boldsymbol k,\lambda}(\eta)
=4\int^\eta d\bar\eta\,
G_{\boldsymbol k}(\eta,\bar\eta)
\sqrt{\mathcal G_\gamma(\bar\eta)}\,
\mathcal S_{\boldsymbol k,\lambda}(\bar\eta),
\end{equation}
where the Green's function  satisfies
\begin{equation}
\left[\partial_\eta^2+
c_T^2k^2
-\frac{(\sqrt{\mathcal G_\gamma})''}
 {\sqrt{\mathcal G_\gamma}}
\right]G_{\boldsymbol k}(\eta,\bar\eta)
=\delta(\eta-\bar\eta).
\end{equation}

For the parameter choice in Eq.~\eqref{choice-CD}, with
\(P=\lambda_0X^2\), one has
\(\mathcal G_\gamma=a^2/8\) and \(c_T^2=1\).  Since \(a''=0\)
during radiation domination, the Green's function reduces to
\begin{equation}
 kG_{\boldsymbol k}(\eta,\bar\eta)
 =\sin(x-\bar x),
\end{equation}
where \(x=k\eta\), \(\bar x=k\bar\eta\).

\subsection{Primordial normalization and scalar transfer functions}

Throughout this subsection, we adopt the parameter
choice~\eqref{choice-CD} and $B_2=0$ during radiation
domination. The background then satisfies
Eq.~\eqref{eq:YB-B2-zero}, with constant $\sigma$.

To express the SIGW spectrum in terms of the primordial curvature
power spectrum, we normalize all scalar
perturbations to the initial superhorizon curvature amplitude,
\begin{equation}
\zeta^{\rm p}_{\boldsymbol k}
\equiv
\lim_{k\eta\to0}
\zeta_{\boldsymbol k}(\eta).
\end{equation}
We introduce dimensionless transfer functions by
writing
\begin{equation}\label{eq:transfer-definitions}
\begin{split}
\zeta_{\boldsymbol k}
&=
T_\zeta(x)\,
\zeta^{\rm p}_{\boldsymbol k},
\qquad
\varphi_{\boldsymbol k}
=
\frac{\bar\chi'}{k}
T_\varphi(x)\,
\zeta^{\rm p}_{\boldsymbol k},
\qquad
\delta\chi_{\boldsymbol k}
=
\frac{\bar\chi'}{k}
T_{\delta\chi}(x)\,
\zeta^{\rm p}_{\boldsymbol k},
\\
A_{\boldsymbol k}
&=
T_A(x)\,
\zeta^{\rm p}_{\boldsymbol k},
\qquad
B_{\boldsymbol k}
=
\frac{1}{k}
T_B(x)\,
\zeta^{\rm p}_{\boldsymbol k}.
\end{split}
\end{equation}
The dependence of the transfer functions on the constant
parameter $\sigma$ is left implicit.
The primordial normalization requires
\begin{equation}\label{NC}
\lim_{x\to0}T_\zeta(x)=1.
\end{equation}
The primordial curvature power spectrum is defined by
\begin{equation}\label{eq:primordial-zeta-spectrum}
\left\langle
\zeta^{\rm p}_{\boldsymbol k}
\zeta^{\rm p}_{\boldsymbol k'}
\right\rangle
=
(2\pi)^3
\delta^{(3)}
\left(
\boldsymbol k+\boldsymbol k'
\right)
\frac{2\pi^2}{k^3}
\mathcal P_\zeta(k).
\end{equation}

\subsubsection{Transfer functions at finite $\sigma$}

Substituting the background relation~\eqref{eq:YB-B2-zero}
into the scalar constraint solutions and using the
definitions~\eqref{eq:transfer-definitions}, we obtain
\begin{equation}
\begin{split}
T_\zeta
={}&
\frac{
6\sigma^2\left[(1+\sigma)^2x^2+3\sigma^2\right]
}{
(1+\sigma)^4x^4
+3\sigma(\sigma+2)(1+\sigma)^2x^2
+18\sigma^3(\sigma-1)
}
\left(
\partial_xT_\varphi
-\frac{2\sigma}{(1+\sigma)x}T_\varphi
\right),
\\[4pt]
T_{\delta\chi}
={}&
T_\varphi+\frac{1+\sigma}{\sigma}xT_\zeta,
\\[4pt]
T_A
={}&
\frac{1+\sigma}{\sigma}x\partial_xT_\zeta
+\frac{2\sigma^2}{(1+\sigma)(\sigma-2)x}T_\varphi
-\frac{2}{\sigma-2}\partial_xT_\varphi
+
\left[
\frac{(1+\sigma)^2x^2}{3\sigma^2(\sigma-2)}
+\frac{2(\sigma^2-\sigma-1)}{\sigma(\sigma-2)}
\right]T_\zeta,
\\[4pt]
T_B
={}&
\frac{6\sigma^2}{(1+\sigma)(\sigma-2)x}
\partial_xT_\varphi
-\frac{6\sigma^2(\sigma+2)}
{(1+\sigma)^2(\sigma-2)x^2}T_\varphi
-
\left[
\frac{1+\sigma}{\sigma-2}x
+\frac{6\sigma}{(1+\sigma)(\sigma-2)x}
\right]T_\zeta.
\end{split}
\end{equation}
The transfer function $T_\varphi$ is obtained numerically by solving
Eq.~\eqref{eq:Phi-eom}, with its coefficients evaluated
using Eq.~\eqref{eq:YB-B2-zero} and $\sigma'=0$.
The primordial normalization condition~\eqref{NC}
fixes the leading superhorizon behavior,
\begin{equation}
T_\varphi(x)
=-\frac{1+\sigma}{\sigma}x+\mathcal O(x^3),
\qquad x\ll1.
\label{eq:finite-sigma-initial-conditions}
\end{equation}
The remaining transfer functions then follow from the constraint
relations above.

\subsubsection{Large-$\lvert\sigma\rvert$ limit}

At leading order in the large-$\lvert\sigma\rvert$ expansion, the transfer functions are
\begin{equation}
\begin{split}
   & T_\zeta
\simeq
\frac{
6(x^2+3)
}{
(x^4+3x^2+18)
}
\left(\partial_x T_\varphi-\frac{2}{x}T_\varphi
\right),\\
&T_{\delta\chi}
\simeq
T_\varphi
+
xT_\zeta,
\\
&T_A
\simeq{}x\partial_xT_\zeta
+\frac{2}{x}T_{\varphi}+2T_\zeta,
\\
&
T_B
\simeq\frac{6}{x}\partial_xT_\varphi
-\frac{6}{x^2}T_{\varphi}
-xT_\zeta
.
\end{split}
\end{equation}
At the same order, the regular solution
in Eq.~\eqref{eq:varphi-sine-solution} gives
\begin{equation}
T_\varphi(x)
=
\mathcal N_\varphi
\sin\left(
\frac{x}{\sqrt3}
\right).
\end{equation}
The primordial normalization condition~\eqref{NC} then yields
\begin{equation}
1
=
\lim_{x\to0}T_\zeta(x)
=
-\frac{\mathcal N_\varphi}{\sqrt3},
\qquad
\mathcal N_\varphi=-\sqrt3.
\end{equation}

\subsection{Modified source function and kernel }

The  power spectrum of SIGWs for each
polarization is given by
\begin{align}
\mathcal{P}_{{\gamma},\lambda}(k,\eta)=4\int_{0}^\infty\mathrm{d}v\int_{|1-v|}^{1+v}\mathrm{d}u
\left[\frac{4v^2-(1+v^2-u^2)^2}{4vu}\right]^2  I^2(k, u,v,x)\mathcal{P}_{\zeta}(u k)\mathcal{P}_{\zeta}(v k),
\end{align}
where $v=p/k$, $u=|{\boldsymbol k}-{\boldsymbol p}|/k$. The kernel $I$ is given by
\begin{equation}\label{IA}
    \begin{split}
I(k, u,v,x)
=\int_{0}^x\mathrm{d}\bar{x}~k~\frac{a(\bar\eta)}{a(\eta)}G_{\boldsymbol k}(x,\bar{x})
f(u,v,\bar x).
    \end{split}
\end{equation}
During radiation domination,
$a(\bar\eta)/a(\eta)=\bar x/x$. For the tensor equation considered here, the Green's function satisfies
$kG_{\boldsymbol k}(\eta,\bar\eta)
=\sin(x-\bar x)$.

To isolate the contribution of the cubic
extrinsic-curvature interaction, we write
the symmetrized source function \footnote{
We define $\boldsymbol p=\boldsymbol k-\boldsymbol q$
and use
$e_\lambda^{ij}p_ip_j=e_\lambda^{ij}q_iq_j$
and
$e_\lambda^{ij}q_ip_j=-e_\lambda^{ij}q_iq_j$.
The convolution kernel is symmetrized according to
$K_{\rm sym}(\boldsymbol q,\boldsymbol p)
=\tfrac12[K(\boldsymbol q,\boldsymbol p)
+K(\boldsymbol p,\boldsymbol q)]$.
} as 
\begin{equation}
f(u,v,x)
=
f_0(u,v,x)
+
f_{c_1}(u,v,x),
\end{equation}
where $f_0$ is the source function for
$c_1^{(0;3,0)}=0$, and $f_{c_1}$ is the additional
term generated by the cubic operator
$c_1^{(0;3,0)}\hat K_i^{j}\hat K_j^{k}\hat K_k^{i}$.

For the parameter choice in Eq.~\eqref{choice-CD} during the
radiation-dominated era, we obtain
\begin{equation}\label{eq:source-function-f0}
\begin{split}
f_0(u,v,x)
={}&
-\left(\frac{1}{4uv}
+\frac{u^2+v^2}{3(1+\sigma)uv}\right)
T_B(ux)T_B(vx)
\\
&+\frac{\sigma}{1+\sigma}
\left[
\frac{1}{v}T_B(vx)\partial_xT_{\zeta}(ux)
+\frac{1}{u}T_B(ux)\partial_xT_{\zeta}(vx)
\right]
\\
&-\left(\frac{\sigma}{1+\sigma}\right)^2
\left[
\frac{1}{ux}T_A(vx)T_B(ux)
+\frac{1}{vx}T_A(ux)T_B(vx)
\right]
\\
&-T_A(vx)T_\zeta(ux)
-T_A(ux)T_\zeta(vx)
-T_\zeta(ux)T_\zeta(vx)
\\
&+\left(\frac{\sigma}{1+\sigma}\right)^2
\frac{2}{uvx^2}
\left[
T_B(vx)T_{\delta\chi}(ux)
+T_B(ux)T_{\delta\chi}(vx)
+T_{\delta\chi}(ux)T_{\delta\chi}(vx)
\right]
\\
&-\frac{1}{x^2}\partial_x
\Bigg[
\frac{3x^2}{4v}T_B(vx)T_\zeta(ux)
+\frac{3x^2}{4u}T_B(ux)T_\zeta(vx)
\\
&\qquad\qquad
-\frac{x^2}{4u}T_A(vx)T_B(ux)
-\frac{x^2}{4v}T_A(ux)T_B(vx)
\Bigg].
\end{split}
\end{equation}
The dependence of $f_0$ on $\sigma$ enters through
both its explicit coefficients and the scalar transfer
functions. The cubic interaction gives
\begin{equation}
f_{c_1}(u,v,x)
=-\frac{k}{a^2}\partial_x
\left[
\frac{ac_1^{(0;3,0)}}{8uv}(u^2+v^2-3)
T_B(ux)T_B(vx)
\right].
\label{eq:source-function-fc1}
\end{equation}
For the numerical analysis, we take $c_1^{(0;3,0)}$ to be
constant and write 
$a(\eta)=a_0\eta$.
We define
\begin{equation}
\mu_s\equiv\frac{c_1^{(0;3,0)}k_s^2}{a_0},
\qquad
\kappa\equiv\frac{k}{k_s},
\end{equation}
where $\mu_s$ is dimensionless and $k_s$ denotes the peak scale of the primordial curvature
spectrum. The corresponding source contribution becomes
\begin{equation}
f_{c_1}(u,v,x)
=
-\frac{\mu_s\kappa^2}{8uvx^2}
(u^2+v^2-3)
\partial_x\!\left[xT_B(ux)T_B(vx)\right].
\end{equation}

The kernel can therefore be written as
\begin{equation}
I(k,u,v,x)
=
I_0(u,v,x)+\mu_s\kappa^2 I_{c_1}(u,v,x).
\end{equation}
Here $I_0$ and $I_{c_1}$ are obtained from
Eq.~\eqref{IA} using $f_0$ and the coefficient of
$\mu_s\kappa^2$ in $f_{c_1}$, respectively. 
After oscillation averaging, the squared kernel becomes
\begin{equation}\label{IA2}
\overline{I^2}
=
\overline{I_0^2}
+2\mu_s\kappa^2\overline{I_0 I_{c_1}}
+\mu_s^2\kappa^4\overline{I_{c_1}^2},
\end{equation}
where the second term is the interference contribution and can
have either sign, whereas the last term is nonnegative.
Their sum determines whether the spectrum is suppressed
or enhanced relative to the result with $\mu_s=0$
at a given scale.

The finite-$\sigma$ effect originates from the additional \(2K^2/[3(1+C_2N)]\) term in Eq.~\eqref{toy-model}, which vanishes as $|\sigma|\to\infty$.  
At finite $\sigma$, the ratio
$Y/\mathcal H^2$ differs from unity according to Eq.~\eqref{eq:YB-B2-zero}, while the background remains radiation dominated.
These corrections also modify the scalar constraints,
the transfer functions, and the explicit coefficients
of $f_0$.
For $\mu_s=0$, the kernel reduces to $I=I_0$. Making its $\sigma$ dependence explicit, we write
\begin{equation}
I_0(\sigma)
=
I_0^{(\infty)}
+
\Delta I_\sigma ,
\end{equation}
where $I_0^{(\infty)}$ denotes the $|\sigma|\to\infty$ limit and
$\Delta I_\sigma$ collects the full finite-$\sigma$ correction.
The sign of the additional $K^2$ coefficient does not determine whether the fractional energy density of SIGWs is enhanced
or suppressed. The effect depends on both the scalar transfer
functions and the explicit source coefficients
through the kernel.

The above analysis identifies two distinct sources of
modifications to SIGWs. Finite-$\sigma$
corrections modify the background
relation, the linear scalar dynamics, and
the source function $f_0$. The cubic
operator
$c_{1}^{(0;3,0)} \hat K_{i}^{j}\hat K_{j}^{k}\hat K_{k}^{i}$ instead contributes an additional term to the source
function and hence to the kernel. It leaves the background and quadratic actions unchanged but modifies the scalar equations
at non-linear order.
The conditions for these corrections to remain
perturbative are discussed in
Appendix~\ref{app:backreaction}.

\subsection{The fractional energy density of SIGWs}
For the parameter choice in Eq.~\eqref{choice-CD},
the quadratic tensor action has the same form and
normalization as in GR on the same FLRW background,
with $c_T^2=1$.
The following calculation is performed with respect to the preferred foliation intrinsic to the SCG formulation. As in the previous study of SIGWs in SCG~\cite{Jiang:2025ysb}, the breaking of temporal reparametrization invariance requires the analysis to be carried out in the unitary gauge.  
 Moreover, the second-order tensor perturbation in the unitary gauge may contain non-radiative contributions in addition to the freely propagating tensor modes. The physical SIGW signal is associated with the latter, and its fractional energy density should be evaluated from the late-time propagating tensor component.

With the polarization convention adopted above,  the
dimensionless tensor power spectrum for each polarization  is defined by
\begin{equation}
\left\langle\gamma_{\boldsymbol k,\lambda}
\gamma_{\boldsymbol k',\lambda'}\right\rangle
=(2\pi)^3\delta_{\lambda\lambda'}
\delta^{(3)}(\boldsymbol k+\boldsymbol k')
\frac{2\pi^2}{k^3}\mathcal P_{\gamma,\lambda}(k,\eta).
\end{equation}
The fractional GW energy density per logarithmic interval in wavenumber is
\begin{equation}
\Omega_{\mathrm{GW}} = \frac{1}{12}\left(\frac{k}{\mathcal{H}}\right)^2 \sum_\lambda \overline{\mathcal{P}_{\gamma,\lambda}},
\end{equation}
where the overline denotes an oscillation average.

In the following, we evaluate the fractional energy density of SIGWs for two
representative forms of the primordial curvature power spectrum: a
monochromatic spectrum and a log-normal spectrum.
All results use the radiation-dominated era
with $B_2=0$.
We examine finite-$\sigma$ corrections with
$\mu_s=0$ and the effect of the cubic coupling
at leading order in the large-$|\sigma|$ expansion.

\subsubsection{Monochromatic primordial spectrum}
\begin{figure}[htp]
\centering
\includegraphics[width=0.46\textwidth]{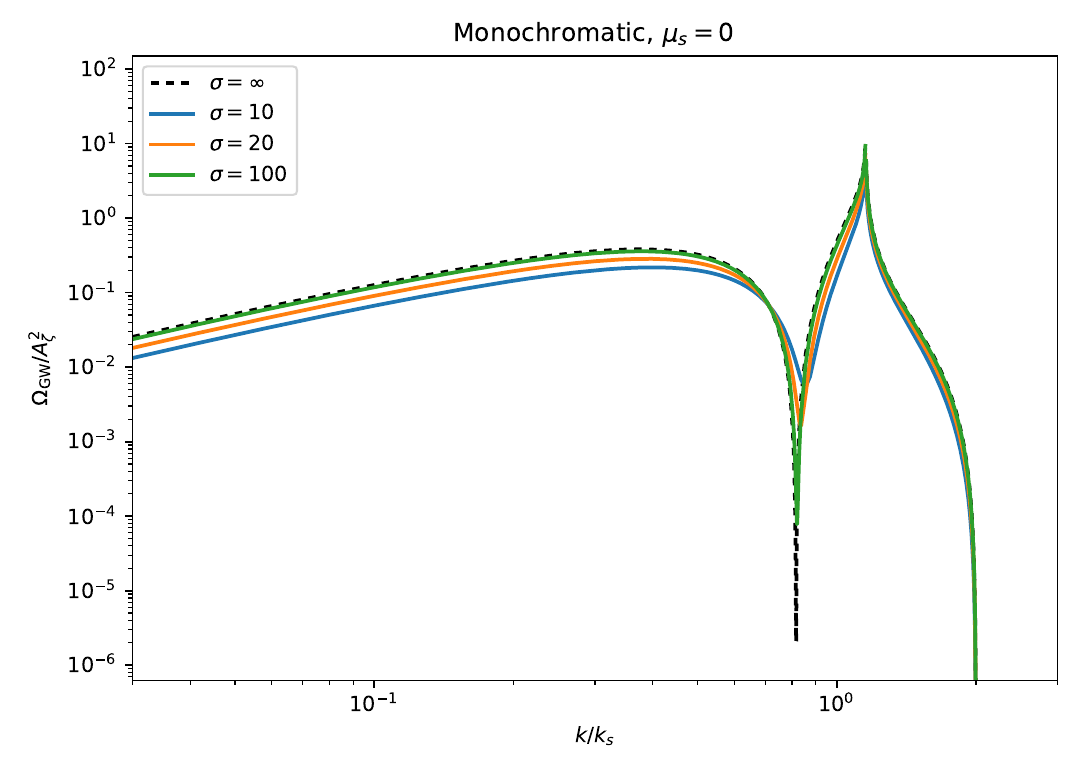}\hspace{0.2cm}
\includegraphics[width=0.46\textwidth]{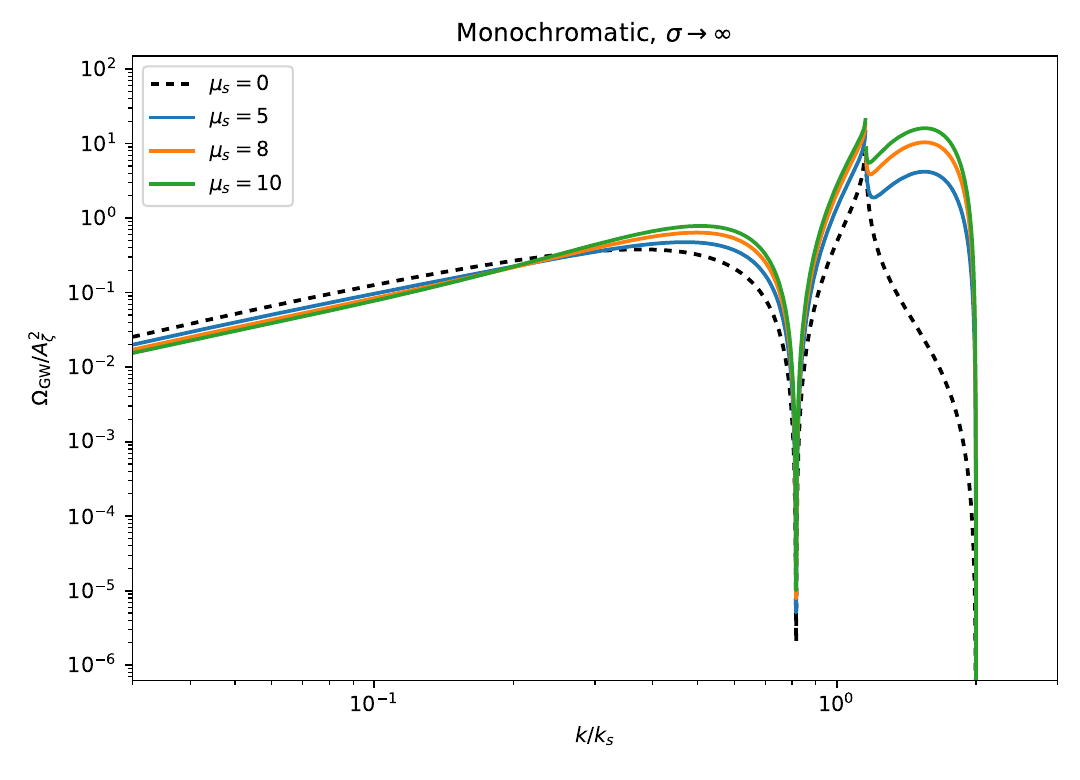}\hspace{0.2cm}
\caption{The fractional energy density $\Omega_{\mathrm{GW}}$ of 
 SIGWs for a monochromatic curvature spectrum. 
 Left: dependence on $\sigma$ with $\mu_s=0$.
Right: dependence on $\mu_s$
in the limit $\sigma\rightarrow\infty$.
The dashed curves correspond to $\sigma\to\infty$
in the left panel and $\mu_s=0$ in the right panel.
}\label{fig1}
\end{figure}

We first consider a monochromatic primordial spectrum for the  curvature perturbation in the form of a Dirac delta function
\begin{equation}
\mathcal P_\zeta(k)=\mathcal A_\zeta \delta\left(\log \frac{k}{k_s}\right),
\end{equation}
where $\mathcal{A}_\zeta$ is the amplitude and $k_s$ denotes the peak scale.
For this spectrum, the momentum convolution is restricted
to $u=v=1/\kappa$, where $\kappa= k/k_s$. 
The fractional energy
density of SIGWs is shown in Fig.~\ref{fig1}.

The left panel of Fig.~\ref{fig1} shows the
dependence of $\Omega_{\mathrm{GW}}$  on $\sigma$ for $\mu_s=0$.
For the positive values of $\sigma$ shown, $\Omega_{\mathrm{GW}}$
is suppressed relative to the
 result in the limit  $\sigma\rightarrow \infty$ over most of the displayed range. The deviation decreases with increasing $\sigma$, and the $\sigma=100$ curve is already close to the asymptotic result, consistent with the $1/\sigma$ expansion discussed above. 

The right panel of Fig.~\ref{fig1} illustrates the effect of the cubic interaction
$c_1^{(0;3,0)}\hat K_{ij}\hat K^{jk}\hat K_k{}^i$ in the limit $\sigma\to\infty$.
For the positive values of $\mu_s$ considered  here, the spectrum is suppressed relative to the $\mu_s=0$ result at small $\kappa$, whereas it is enhanced at larger $\kappa$, where an additional broad maximum develops. This behavior can be understood from the kernel
decomposition in Eq.~\eqref{IA2}, which separates
the correction into an interference term and a
nonnegative term quadratic in $\mu_s$.
Using $u=v=1/\kappa$, the momentum factor
in the cubic source  reduces to
\begin{equation}
\frac{u^2+v^2-3}{uv}=2-3\kappa^2,
\end{equation}
which changes sign at $\kappa=\sqrt{2/3}$.
At small $\kappa$,  the suppression
implies that the interference term is negative
and exceeds the quadratic contribution in magnitude.
At larger $\kappa$, the enhancement implies
that the sum of the two contributions becomes positive. The additional broad maximum becomes more pronounced
as $\mu_s$ increases, demonstrating that the cubic
interaction modifies the spectral shape rather than
merely rescaling the amplitude.

\subsubsection{Log-normal primordial spectrum}
\begin{figure}[htp]
\centering
\includegraphics[width=0.46\textwidth]
{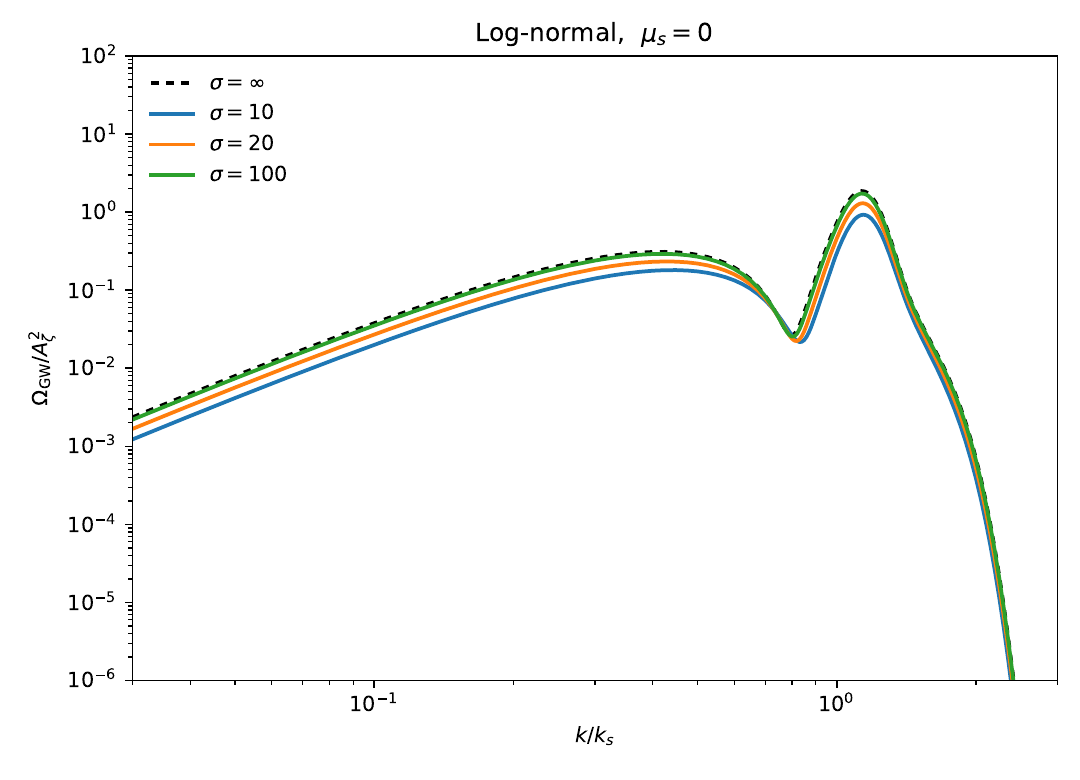}\hspace{0.2cm}
\includegraphics[width=0.46\textwidth]
{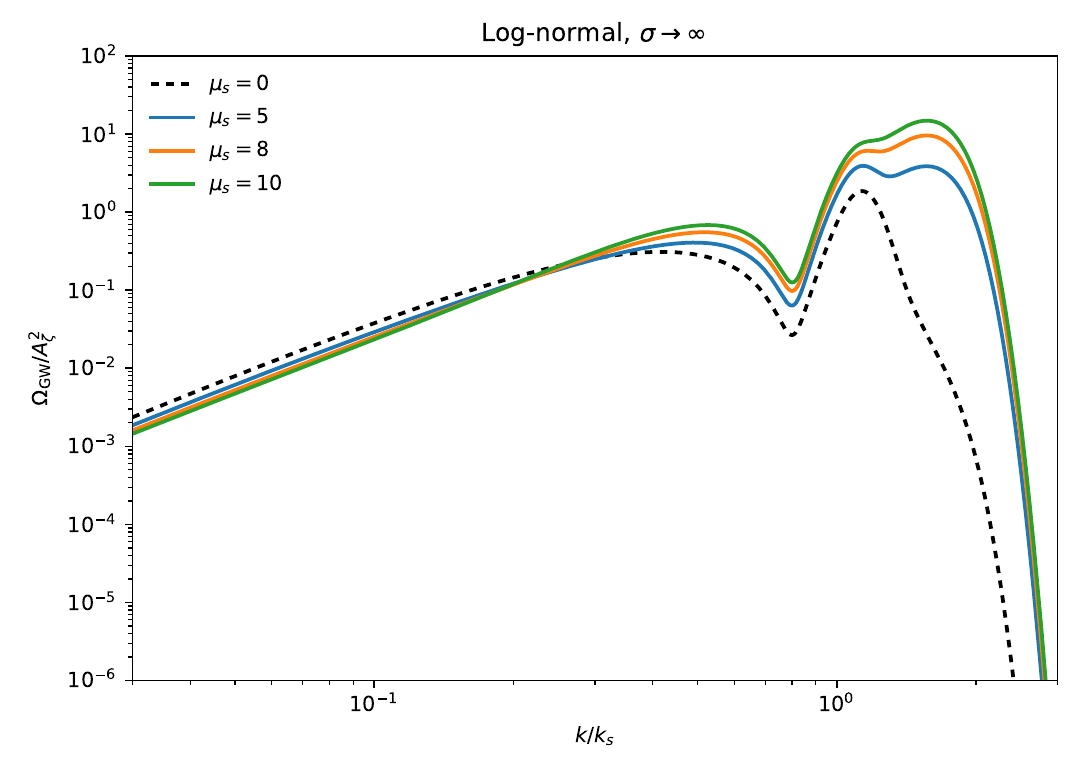}\hspace{0.2cm}
\caption{The fractional energy density $\Omega_{\mathrm{GW}}$ of 
 SIGWs for a log-normal primordial spectrum with $\Delta_{\zeta}=0.1$.
Left: dependence on $\sigma$ with $\mu_s=0$.
Right: dependence on
$\mu_s$
in the limit $\sigma\rightarrow\infty$.
The dashed curves correspond to $\sigma\to\infty$
in the left panel and $\mu_s=0$ in the right panel.
}\label{fig2}
\end{figure}

We next consider a log-normal primordial curvature spectrum with
a finite width,
\begin{equation}
\mathcal P_\zeta(k)= \frac{\mathcal A_\zeta}{\sqrt{2 \pi} \Delta_\zeta} \exp \left[-\frac{\log ^2\left(k / k_s\right)}{2 \Delta_{\zeta}^2}\right],
\end{equation}
where $\Delta_{\zeta}$ is the dimensionless width.  We take
$\Delta_{\zeta}=0.1$ in the numerical
calculation. The finite width of the primordial spectrum allows the momentum convolution to sample a range of $u$
and $v$, rather than only $u=v=1/\kappa$. The corresponding results are shown in
Fig.~\ref{fig2}.

The left panel of Fig.~\ref{fig2} exhibits a dependence
on $\sigma$  similar to that in the monochromatic case. For the positive values of $\sigma$ shown, the energy density of SIGWs 
is suppressed relative to the  result in the limit  $\sigma\rightarrow \infty$ over most of
the displayed range, and the deviation decreases as $\sigma$
increases. In particular, the $\sigma=100$ curve is already close
to the asymptotic result. Compared with the monochromatic case, the features of the fractional energy density are smoother, reflecting the finite width
of the primordial spectrum and the integration
over a distribution of scalar momenta.

The right panel of Fig.~\ref{fig2} shows the effect
of the cubic interaction. For the positive values of
$\mu_s$ considered here, the spectrum is  suppressed at small $\kappa$ and 
enhanced at large $\kappa$, with a broad
maximum becoming more pronounced as $\mu_s$ increases. 
Compared with the monochromatic case, the convolution
over a range of scalar momenta, rather than being restricted to  $u=v=1/\kappa$,
smooths the sharp peak.
The factor $(u^2+v^2-3)/(uv)$ takes both positive
and negative values within the integration domain,
and its effect must be evaluated using the full kernel.
At larger $\kappa$, the observed enhancement indicates
that the sum of the interference and quadratic
contributions in Eq.~\eqref{IA2} is positive after
integration over the scalar momenta.

\section{Conclusion}\label{Sec.7}
In this work, we studied SIGWs in 2-DOF spatially covariant gravity minimally
coupled to a $k$-essence scalar field  with Lagrangian $P(\chi,X)$. For general coefficient functions $B_i$,
$C_i$, and $D_i$ satisfying the defining conditions
of the model, we derived the quadratic actions
for tensor and scalar perturbations, together
with the cubic action involving two scalar
perturbations and one tensor perturbation.
After integrating out the lapse and scalar shift
perturbations, we showed that the kinetic matrix
of the coupled matter and metric scalar perturbations
is degenerate. Provided that $D_{\chi'\chi'}\neq0$,
the scalar kinetic terms can therefore be expressed
in terms of a single scalar combination.

We then specialized to a radiation-dominated era with
$P(\chi,X)=\lambda_0X^2$  and considered  the parameter
choice~\eqref{choice-CD},  for which the quadratic
tensor action coincides with that of GR.
For $B_2=0$, the background equations require $\sigma$
to be constant and fix $B_1/\bar N$.
We derived the scalar perturbation equations at finite
$\sigma$ and obtained analytical solutions in the limit
$|\sigma|\rightarrow\infty$.
Using the corresponding scalar transfer functions,
we evaluated the source function and kernel of SIGWs.

We separately examined the effects of  finite $\sigma$ and the cubic interaction
$c_1^{(0;3,0)}\hat K_i^{j}\hat K_j^{k}\hat K_k^{i}$. Finite $\sigma$ modifies the scalar evolution and
the explicit coefficients of the source function. The cubic interaction leaves the background and quadratic
perturbation actions unchanged but provides an
additional contribution to the source function and hence the kernel. The resulting correction to the energy density
contains an interference term and a nonnegative term quadratic in the coupling.

We evaluated the fractional energy density
of SIGWs for monochromatic and log-normal
primordial curvature power spectra. For $\mu_s=0$ and the positive values of
$\sigma$ studied, the spectrum is suppressed over
a broad range of scales relative to the result
in the limit $\sigma\to\infty$.
The deviation decreases as $\sigma$ increases.
In the limit $\sigma\to\infty$, the positive values
of $\mu_s$ studied produce suppression at small
$\kappa$ and enhancement at larger $\kappa$,
with an additional broad maximum becoming more
pronounced as $\mu_s$ increases. 
The suppression arises when the negative interference
contribution exceeds the positive quadratic
term in magnitude, whereas enhancement occurs when their sum is positive.

Our results show that the fractional energy density of SIGWs can retain
distinctive features of the underlying gravity model
even when the quadratic tensor action coincides
with that of GR.
These features arise from modified scalar dynamics
and interactions that first appear at cubic order. Our numerical analysis focused on a radiation-dominated era with the parameter choice~\eqref{choice-CD},  $B_2=0$,
and a constant cubic coupling $c_1^{(0;3,0)}$.
More general backgrounds and parameter choices 
with $B_2\neq0$, including those admitting
time-dependent $\sigma$, as well as a time-dependent
cubic coupling, are left for future work.

\begin{acknowledgments}
We thank Fengge Zhang for helpful discussions. This work was supported in part by the National Natural Science Foundation of China (NSFC) under Grants No. 12547125 (J.X.F.), No. 12547120 (Y.Y.), Nos. 12547110, 12475065, and 12447101 (Q.G.H.), and Nos. 12475068 and 11975020 (X.G.), and by the Guangdong Basic and Applied Basic Research Foundation under Grant No. 2025A1515012977 (X.G.).

\end{acknowledgments}

\appendix

\section{General reduced scalar coefficients}\label{app:QD}

\subsection{The coefficients $Q_{nn}$}
The coefficients $Q_{nn}$ in the quadratic scalar action \eqref{QC1} are
\begin{equation}  
\begin{split}
Q_{AA}
&=\frac{3\mathcal H^2 C_1C_2^2\bar N^3}{2a^2(1+C_2\bar N)^3}
+\frac{9\mathcal H^3 }{2a^3}\frac{C^2_2D_1\bar N^4}{(1+C_2\bar N)^4}
+\bar{X}\bar{P}_X+2\bar{X}^2\bar{P}_{XX}\\
Q_{A\zeta}&=\frac{2k^2}{a^2}C_4,\\
Q_{A\zeta'}&=-\frac{3\mathcal H}{a^2}\left(\frac{C_1C_2\bar{N}^2}{(1+C_2\bar{N})^2}
+\frac{3\mathcal HC_2 D_1\bar{N}^3}{a(1+C_2\bar{N})^3}
\right),\\
Q_{AB}&=-\frac{k^2\mathcal H}{a^2 }\left(\frac{C_1C_2\bar{N}^2}{(1+C_2\bar{N})^2}
+\frac{3\mathcal HC_2 D_1\bar{N}^3}{a(1+C_2\bar{N})^3}
\right),\\
\end{split}
\end{equation}
\begin{equation}  
\begin{split}
Q_{B\zeta}&=
\frac{2k^4C_4D_2}{a^3C_1C_2(C_5-2C_2\bar N)},\\
Q_{B\zeta'}&=\frac{k^2}{a^2}\left(\frac{C_1\bar{N}}{(1+C_2\bar{N})}
+\frac{3\mathcal H D_1\bar{N}^2}{a(1+C_2\bar{N})^2}
\right), \\
Q_{BB}&
=\frac{k^4C_1\bar N(C_5+2)}{6a^2(1+C_2\bar N)(C_5-2C_2\bar N)}\\
&\quad
+\frac{k^4\mathcal HD_2\bar N^2}{a^3(1+C_2\bar N)(C_5-2C_2\bar N)^2}
+\frac{k^4\mathcal HD_1\bar N^2(C_5+2)}{2a^3(1+C_2\bar N)^2(C_5-2C_2\bar N)}.
\end{split}
\end{equation}
\begin{equation}
\begin{split}
  &Q_{\zeta'\zeta'}=\frac{3}{2a^2(1+C_2\bar N)}
\left[C_1 \bar N+\frac{3\mathcal H D_1\bar N^2}{a(1+C_2\bar N)}\right],\\[3pt]
   &Q_{\zeta'\zeta}=\frac{2k^2}{a^3}D_3,\\
& Q_{\zeta\zeta}=
\frac{k^2}{a^2}\left(\frac{C_3}{\bar{N}}+\frac{\mathcal{H}D_3}{a}+C_4\right), 
\end{split}
\end{equation}
and
\begin{equation}
    \begin{aligned}
&Q_{\chi'\chi'}=\frac{1}{2a^2}(\bar{P}_X+2\bar{X}\bar{P}_{XX}),
&Q_{\chi'\chi}&=\frac{\bar{\chi}'\bar{P}_{\chi X}}{a^2},  \\
&Q_{\chi\chi}=\frac{\bar{P}_{\chi\chi}}{2}-\frac{k^2}{2a^2}\bar{P}_X, 
&Q_{\chi'A}&=-\frac{\bar{\chi}'}{a^2}(\bar{P}_X+2\bar{X}\bar{P}_{XX}), \\
&Q_{\chi B}=-\frac{k^2}{a^2}\bar{\chi}'\bar{P}_X,
&Q_{\chi A}&=\bar{P}_{\chi}-2\bar{X}\bar{P}_{\chi X},\\
&Q_{\chi'\zeta}=\frac{3\bar{\chi}'\bar{P}_X}{a^2}, 
&Q_{\chi \zeta}&=3\bar{P}_{\chi},
    \end{aligned}
\end{equation}
where we have used the background equations \eqref{L1-eom} to simplify the expressions.

\subsection{The coefficients $D_{nn}$}
The coefficients $D_{nn}$ in the quadratic scalar action \eqref{QC2} are
\begin{equation}\label{general-kinetic-D}
\begin{split}
        D_{\zeta'\zeta'}&=Q_{\zeta'\zeta'}-\frac{Q_{BB}Q_{A\zeta'}^2-Q_{AB}Q_{A\zeta'}Q_{B\zeta'}+Q_{AA}Q_{B\zeta'}^2}{\Delta}\\
        D_{\zeta'\zeta}&=Q_{\zeta'\zeta}-\frac{2Q_{BB}Q_{A\zeta'}Q_{A\zeta}-Q_{AB}(Q_{A\zeta'}Q_{B\zeta}+Q_{A\zeta}Q_{B\zeta'})+2Q_{AA}Q_{B\zeta'}Q_{B\zeta}}{\Delta}\\
          D_{\zeta\zeta}&=Q_{\zeta\zeta}-\frac{Q_{BB}Q^2_{A\zeta}-Q_{AB}Q_{A\zeta}Q_{B\zeta}+Q_{AA}Q^2_{B\zeta}}{\Delta}\\
D_{\chi'\chi'}
&=
Q_{\chi'\chi'}
-\frac{Q_{BB}Q_{\chi'A}^{\,2}}{\Delta},
\\[4pt]
D_{\chi'\chi}
&=
Q_{\chi'\chi}
-\frac{
2Q_{BB}Q_{\chi'A}Q_{\chi A}
-Q_{AB}Q_{\chi'A}Q_{\chi B}
}{\Delta},
\\[4pt]
D_{\chi\chi}
&=
Q_{\chi\chi}
-\frac{
Q_{BB}Q_{\chi A}^{\,2}
-Q_{AB}Q_{\chi A}Q_{\chi B}
+Q_{AA}Q_{\chi B}^{\,2}
}{\Delta}.
\\
D_{\zeta'\chi'}
&=
-\frac{
2Q_{BB}Q_{A\zeta'}Q_{\chi'A}
-Q_{AB}Q_{B\zeta'}Q_{\chi'A}
}{\Delta},
\\[4pt]
D_{\zeta\chi'}
&=
Q_{\chi'\zeta}
-\frac{
2Q_{BB}Q_{A\zeta}Q_{\chi'A}
-Q_{AB}Q_{B\zeta}Q_{\chi'A}
}{\Delta},
\\[4pt]
D_{\zeta'\chi}
&=
-\frac{
2Q_{BB}Q_{A\zeta'}Q_{\chi A}
-Q_{AB}
\left(
Q_{A\zeta'}Q_{\chi B}
+Q_{B\zeta'}Q_{\chi A}
\right)
+2Q_{AA}Q_{B\zeta'}Q_{\chi B}
}{\Delta},
\\[4pt]
D_{\zeta\chi}
&=
Q_{\chi\zeta}
-\frac{
2Q_{BB}Q_{A\zeta}Q_{\chi A}
-Q_{AB}
\left(
Q_{A\zeta}Q_{\chi B}
+Q_{B\zeta}Q_{\chi A}
\right)
+2Q_{AA}Q_{B\zeta}Q_{\chi B}
}{\Delta},
\end{split}
\end{equation}
where 
\begin{equation}\label{delta}
    \Delta \equiv 4Q_{AA}Q_{BB}-Q^2_{AB}.
\end{equation}

\subsection{The coefficients $\mathcal D_{nn}$ }
The coefficients $\mathcal D_{nn}$ in the quadratic scalar action \eqref{QC3}
are
\begin{equation}
    \begin{aligned}
\mathcal D_{\varphi'\varphi'}&=D_{\chi'\chi'},
&
\mathcal D_{\varphi\varphi'}&=D_{\chi'\chi},
&
\mathcal D_{\varphi\varphi}&=D_{\chi\chi},
\\[3pt]
\mathcal D_{\zeta\varphi'}
&=2r'D_{\chi'\chi'}+rD_{\chi'\chi}+D_{\zeta\chi'},
&
\mathcal D_{\zeta'\varphi}
&=rD_{\chi'\chi}+D_{\zeta'\chi},
\\[3pt]
\mathcal D_{\zeta\varphi}
&=r'D_{\chi'\chi}+2rD_{\chi\chi}+D_{\zeta\chi},
\end{aligned}
\end{equation}
and
\begin{equation}
\begin{split}
\mathcal D_{\zeta'\zeta}
={}&D_{\zeta'\zeta}
+2rr'D_{\chi'\chi'}
+r'D_{\zeta'\chi'}
+r^2D_{\chi'\chi}
+rD_{\zeta\chi'}
+rD_{\zeta'\chi},
\\[3pt]
\mathcal D_{\zeta\zeta}
={}&D_{\zeta\zeta}
+r'^2D_{\chi'\chi'}
+rr'D_{\chi'\chi}
+r'D_{\zeta\chi'}
+r^2D_{\chi\chi}
+rD_{\zeta\chi}.
\end{split}
\end{equation}
With  Eqs.~\eqref{detK} and \eqref{general-Phi}, we have
\begin{equation}
\begin{split}
    &\mathcal D_{\zeta'\zeta'}
=D_{\zeta'\zeta'}+rD_{\zeta'\chi'}+r^2D_{\chi'\chi'}=0,
\\
&\mathcal D_{\zeta'\varphi'}
=2rD_{\chi'\chi'}+D_{\zeta'\chi'}=0.
\end{split}
\end{equation}

\section{Degeneracy of the kinetic matrix for general parameters}
\label{app:general-degeneracy}

We now show that the kinetic matrix of the coupled scalar system is
degenerate without specifying the explicit forms of the coefficient
functions \(C_i\), \(D_i\), and \(B_i\) within 
\(S^{\mathrm{A}}\).

For convenience, we separate the purely gravitational contribution
to \(Q_{AA}\) from the matter contribution by writing
\begin{equation}
Q_{AA}
=
Q_{AA}^{(g)}+\Sigma,
\end{equation}
where
\begin{equation}
Q_{AA}^{(g)}
=
\frac{3\mathcal H^2}{2a^2}
\frac{C_1C_2^2\bar N^3}
{\left(1+C_2\bar N\right)^3}+\frac{9\mathcal H^3 }{2a^3}\frac{C^2_2D_1\bar N^4}{(1+C_2\bar N)^4},
\qquad\quad
\Sigma
\equiv
\bar X
\left(
\bar P_X+2\bar X\bar P_{XX}
\right).
\end{equation}
The matter kinetic coefficients are
\begin{equation}
Q_{\chi'\chi'}
=
\frac{1}{2a^2}
\left(
\bar P_X+2\bar X\bar P_{XX}
\right),
\qquad\qquad
Q_{\chi'A}
=
-\frac{\bar\chi'}{a^2}
\left(
\bar P_X+2\bar X\bar P_{XX}
\right).
\end{equation}
Using $\bar\chi'^2=2a^2\bar X$,
one obtains the identity
\begin{equation}
Q_{\chi'A}^2
=
4 Q_{\chi'\chi'}\Sigma.
\label{matter-degeneracy-relation}
\end{equation}

The purely gravitational counterpart of $\Delta$ in Eq.~\eqref{delta}, is defined as
\begin{equation}
\Delta_g
\equiv
4Q_{AA}^{(g)}Q_{BB}-Q_{AB}^2.
\end{equation}
These quantities satisfy 
\begin{equation}
\Delta
=
\Delta_g+4Q_{BB}\Sigma.
\label{Delta-relation}
\end{equation}
With the convention
\begin{equation}
\mathbf K=
\begin{pmatrix}
2D_{\zeta'\zeta'} & D_{\zeta'\chi'}\\
D_{\zeta'\chi'} & 2D_{\chi'\chi'}
\end{pmatrix},
\end{equation}
the determinant is
\begin{equation}
\det\mathbf K
=
4D_{\zeta'\zeta'}D_{\chi'\chi'}
-D_{\zeta'\chi'}^2.
\end{equation}
By using Eq.~\eqref{matter-degeneracy-relation} and
\eqref{Delta-relation}, the  coefficient $D_{\chi'\chi'}$
can be simplified as
\begin{equation}
\begin{split}
D_{\chi'\chi'}
=
Q_{\chi'\chi'}
\left(
1-\frac{4 Q_{BB}\Sigma}{\Delta}
\right)
=
Q_{\chi'\chi'}
\frac{\Delta_g}{\Delta}.
\end{split}
\label{D-chichi-general}
\end{equation}
Substituting Eq.~\eqref{general-kinetic-D} into  the determinant $\det\mathbf K$
and using
Eqs.~\eqref{matter-degeneracy-relation}--%
\eqref{D-chichi-general}, we find
\begin{equation}\label{detK-general-intermediate}
\begin{split}
\det\mathbf K
&=
\frac{4Q_{\chi'\chi'}}{\Delta}
\Big[
Q_{\zeta'\zeta'}\Delta_g
-Q_{BB}Q_{A\zeta'}^2
+Q_{AB}Q_{A\zeta'}Q_{B\zeta'}
-Q_{AA}^{(g)}Q_{B\zeta'}^2
\Big]
\end{split}
\end{equation}

To evaluate the gravitational coefficients,
we introduce
\begin{equation}
F\equiv
\frac{C_1\bar N}{1+C_2\bar N}
+
\frac{3\mathcal H D_1\bar N^2}
{a(1+C_2\bar N)^2},
\qquad
\Theta\equiv
\frac{\mathcal H C_2\bar N}
{1+C_2\bar N}.
\end{equation}
The relevant coefficients can then be written as
\begin{equation}
\begin{aligned}
Q_{\zeta'\zeta'}&=\frac{3F}{2a^2},
&
Q_{A\zeta'}&=-\frac{3\Theta F}{a^2},
&
Q_{AA}^{(g)}&=\frac{3\Theta^2F}{2a^2},
&
Q_{B\zeta'}&=\frac{k^2F}{a^2},
&
Q_{AB}&=-\frac{k^2\Theta F}{a^2}.
\end{aligned}
\label{general-gravity-kinetic-Q}
\end{equation}
These expressions retain the full $D_1$ dependence.
They satisfy
\begin{equation}
\begin{split}
4Q_{\zeta'\zeta'}Q_{AA}^{(g)}
-Q_{A\zeta'}^2
&=
\frac{9\Theta^2F^2}{a^4}
-\frac{9\Theta^2F^2}{a^4}
=0,\\
-Q_{\zeta'\zeta'}Q_{AB}^2
+Q_{AB}Q_{A\zeta'}Q_{B\zeta'}
-Q_{AA}^{(g)}Q_{B\zeta'}^2
&=
\left(-\frac32+3-\frac32\right)
\frac{k^4\Theta^2F^3}{a^6}
=0.
\end{split}
\label{gravity-identity-2}
\end{equation}
Consequently, the expression in brackets
in Eq.~\eqref{detK-general-intermediate}
reduces to  \begin{equation}
\begin{split}
&
Q_{\zeta'\zeta'}\Delta_g
-Q_{BB}Q_{A\zeta'}^2
+Q_{AB}Q_{A\zeta'}Q_{B\zeta'}
-Q_{AA}^{(g)}Q_{B\zeta'}^2
\\
={}&
Q_{BB}
\left(
4Q_{\zeta'\zeta'}Q_{AA}^{(g)}
-Q_{A\zeta'}^2
\right)
-Q_{\zeta'\zeta'}Q_{AB}^2
+Q_{AB}Q_{A\zeta'}Q_{B\zeta'}
-Q_{AA}^{(g)}Q_{B\zeta'}^2
\\
={}&0.
\end{split}
\end{equation}
Thus,
\begin{equation}
\det\mathbf K=0
\end{equation}
for general coefficient functions \(C_i\), \(D_i\), and \(B_i\)
within  \(S^{\mathrm{A}}\), provided  $\Delta\neq0$.

\section{ Cubic action $S_{\gamma ss}^{(3)}$}\label{Cubic-action}
The cubic action for scalar--scalar--tensor interactions is
\begin{equation}
\begin{split}
S_{ \gamma ss}^{(3)}
=&
\int d\eta \int \frac{d^3 k}{(2\pi)^3}  \frac{d^3 q}{(2\pi)^3}  
~\Bigg[\gamma^{ij}(-\boldsymbol{k})q_iq_j\Big(  
\mathcal F_{BB}  B_{\boldsymbol{q}}B_{\boldsymbol{k-q}}
+  \mathcal F_{B\zeta} 
B_{\boldsymbol{q}}\zeta_{\boldsymbol{k-q}}
+  \mathcal F_{B\zeta'}B_{\boldsymbol{q}} \zeta'_{\boldsymbol{k-q}}
\\
&
+  \mathcal F_{AB}  A_{\boldsymbol{q}}B_{\boldsymbol{k-q}}
+  \mathcal F_{A\zeta} A_{\boldsymbol{q}} \zeta_{\boldsymbol{k-q}}
+  \mathcal F_{\zeta\zeta}\zeta_{\boldsymbol{q}}\zeta_{\boldsymbol{k-q}}
+ \mathcal F_{\zeta'\zeta}\zeta'_{\boldsymbol{q}}\zeta_{\boldsymbol{k-q}}
\\
&
+\mathcal F_{B\chi} B_{\boldsymbol{q}}\delta\chi_{\boldsymbol{k-q}} 
+     \mathcal F_{\chi\chi} \delta\chi_{\boldsymbol{q}} \delta\chi_{\boldsymbol{k-q}} \Big)\\
&+  {\gamma'}^{ij}(-\boldsymbol{k})q_iq_j
\Big(
\mathcal G_{BB} B_{\boldsymbol{q}}B_{\boldsymbol{k-q}}
+\mathcal G_{B\zeta} B_{\boldsymbol{q}}\zeta_{\boldsymbol{k-q}}
+\mathcal G_{B\zeta'} B_{\boldsymbol{q}}\zeta'_{\boldsymbol{k-q}}\\
&
+\mathcal G_{AB} A_{\boldsymbol{q}}B_{\boldsymbol{k-q}}
+\mathcal G_{A\zeta} A_{\boldsymbol{q}}\zeta_{\boldsymbol{k-q}}
+\mathcal G_{\zeta\zeta}\zeta_{\boldsymbol{q}}\zeta_{\boldsymbol{k-q}} 
\Big)\Bigg]
\end{split}
\end{equation}
where
\begin{equation}
   \begin{split}
        &\mathcal F_{BB}=
        -\frac{a^2C_1\bar N}{4(C_5-2C_2\bar N)}k^2
-\frac{a^2C_1\bar N(C_5+2)}{6(1+C_2\bar N)(C_5-2C_2\bar N)}\left(q^2+|\boldsymbol{k-q}|^2\right)\\[5pt]
&\qquad\quad
-\frac{3a\mathcal H\bar N^2[D_1(C_5-2C_2\bar N)+D_2]}{4(1+C_2\bar N)(C_5-2C_2\bar N)^2}k^2\\[5pt]
&\qquad\quad
-\frac{a\mathcal H\bar N^2\big[2(1+C_2\bar N)D_2+D_1(C_5-2C_2\bar N)(C_5+2)\big]}{2(1+C_2\bar N)^2(C_5-2C_2\bar N)^2}\left(q^2+|\boldsymbol{k-q}|^2\right),\\[5pt]
          & \mathcal F_{B\zeta}=-\frac{2aC_4D_2}{C_1C_2(C_5-2C_2\bar N)}\left(q^2+|\boldsymbol{k-q}|^2\right),
        \\[5pt]
   & \mathcal F_{B\zeta'}=-\frac{a^2C_1\bar{N}}{1+C_2\bar{N}}
   -\frac{3a\mathcal H D_1\bar N^2}{(1+C_2\bar N)^2}
      \end{split}
\end{equation}
   \begin{equation}
   \begin{split}
   &\mathcal F_{AB}=
    \frac{a^2\mathcal H C_1C_2\bar{N}^2}{(1+C_2\bar{N})^2}    
    +\frac{3a\mathcal H^2 C_2D_1\bar N^3}{(1+C_2\bar N)^3}
+\frac{3k^2aC_4D_2\bar N}{C_1(C_5-2C_2\bar N)^2},
  \\
  &
  \mathcal F_{A\zeta}=-2a^2C_4
    \\
    &\mathcal F_{\zeta\zeta} =-a^2\left(\frac{C_3}{\bar{N}}+ C_4+\frac{\mathcal H D_3}{a}\right)
   \\
   &\mathcal F_{\zeta'\zeta} =-2aD_3
   \\
  & \mathcal F_{B\chi}  = a^2 \bar{\chi}'\bar{P}_{X}
  \\
  &
\mathcal F_{\chi\chi} =\frac{a^2}{2}\bar{P}_{X}
   \end{split}
\end{equation}
and
\begin{equation}
    \begin{split}
     &\mathcal G_{BB}=-\frac{a}{4}c_1^{(0 ; 3,0)}\left(q^2+|\boldsymbol{k-q}|^2+3\boldsymbol{q}\cdot (\boldsymbol{k-q})\right)
    +\frac{a\bar N^2[D_1(C_5-2C_2\bar N)+D_2]}{4(1+C_2\bar N)(C_5-2C_2\bar N)^2}\left(q^2+|\boldsymbol{k-q}|^2 \right)
        \\[5pt]
       & \mathcal G_{B\zeta}=\frac{3a^2C_1\bar{N}}{2(C_5-2C_2\bar{N})}
       +\frac{9a\mathcal H\bar N^2[D_1(C_5-2C_2\bar N)+D_2]}{2(1+C_2\bar N)(C_5-2C_2\bar N)^2}
      \\[5pt]
   & \mathcal G_{B\zeta'}=\frac{3a\bar N^2[D_1(C_5-2C_2\bar N)+D_2]}{2(1+C_2\bar N)(C_5-2C_2\bar N)^2}\\[5pt]
   & \mathcal G_{AB}=\frac{a^2C_1C_2\bar{N}^2}{(C_5-2C_2\bar{N})^2}
  +\frac{3a\mathcal H C_2D_2\bar N^3(4-C_5+6C_2\bar N)}{2(1+C_2\bar N)^2(C_5-2C_2\bar N)^3}
+\frac{3a\mathcal H C_2D_1\bar N^3(2-C_5+4C_2\bar N)}{2(1+C_2\bar N)^2(C_5-2C_2\bar N)^2}
  \\[5pt]
   & \mathcal G_{A\zeta}=
    \frac{3aC_4D_2\bar N}{C_1(C_5-2C_2\bar N)^2}
        \\[5pt]
  &  \mathcal G_{\zeta\zeta}=-aD_3
  +\frac{3aC_4D_2}{C_1C_2(C_5-2C_2\bar N)}.
    \end{split}
\end{equation}

\section{Non-linear backreaction}\label{app:backreaction}

We estimate the regime of validity of perturbation theory by
comparing the scalar response generated by the
$c_1^{(0;3,0)}$ interaction with the linear solution.
The relevant higher-derivative cubic terms are
\begin{equation}
\begin{split}
\left.S_{c_1}^{(3)}\right|_{\rm h.d.}
=
\int d\eta\,d^3x\,
a\,c_1^{(0;3,0)}
\Bigg[
&
\frac12(\partial^2B)
\partial_i\partial_jB\,\partial^i\partial^jB
-\frac12
\partial_i\partial_jB\,
\partial^j\partial_kB\,
\partial^k\partial^iB
\\
&
-\frac19(\partial^2B)^3
-\frac12\gamma'^{ij}
\partial_i\partial_jB\,\partial^2B
+\frac34\gamma'^{ij}
\partial_i\partial_\ell B\,
\partial_j\partial^\ell B
\Bigg].
\end{split}
\end{equation}
The cubic scalar and scalar--scalar--tensor interactions modify
the scalar equation of motion at non-linear order.

\subsection{Finite $\sigma$}
At finite $\sigma$, integrating the mixed term in
Eq.~\eqref{QC4} by parts gives
\begin{equation}
S_\varphi^{(2)}
=\int d\eta\,\frac{d^3k}{(2\pi)^3}\,
K_k\left(
\varphi_k'\varphi_{-k}'
-\Omega_k^2\varphi_k\varphi_{-k}
\right),
\label{eq:backreaction-S2-finite}
\end{equation}
where
\begin{equation}
K_k
\equiv a^4\widehat{\mathcal D}_{\varphi'\varphi'}
=
\frac{6a^2\mathcal H^2Y\sigma^3\mathcal P_\sigma}
{\bar\chi'^2\mathcal L_\sigma}.
\label{eq:backreaction-K-finite}
\end{equation}
Using the background identities, we have $K_k'/K_k=\Gamma_k$.
Variation of the quadratic action therefore gives
\begin{equation}
\frac{\delta S_\varphi^{(2)}}{\delta\varphi_{-k}(\eta)}
=-2K_k\left(
\varphi_k''+\Gamma_k\varphi_k'
+\Omega_k^2\varphi_k
\right).
\label{eq:backreaction-variation-S2}
\end{equation}
At finite $\sigma$, the linear scalar evolution is governed by
Eq.~\eqref{eq:Phi-eom}, with the full coefficients
$\Gamma_k$ and $\Omega_k^2$.

Substituting the algebraic curvature solution into the shift
constraint gives
\begin{equation}
B_k=\beta_{0,k}\varphi_k+\beta_{1,k}\varphi_k',
\end{equation}
where
\begin{equation}
\begin{split}
\beta_{0,k}
=
\frac{6Y}{\bar\chi'k^2}
\left[
1-
\frac{36\mathcal H^2\sigma^3(1+\sigma)^2Y^2}
{\mathcal L_\sigma}
\right],
\qquad\qquad
\beta_{1,k}
=
\frac{108\mathcal H Y^3\sigma^2(1+\sigma)^3}
{\bar\chi'k^2\mathcal L_\sigma}.
\end{split}
\end{equation}
These expressions use the background relation
\eqref{eq:YB-B2-zero} and the linear algebraic constraints.

For comparable characteristic momenta, we neglect numerical
coefficients and angular kernels while retaining the full
finite-$\sigma$ shift solution. Combining the quadratic and
cubic variations gives
\begin{equation}
\varphi_k''+\Gamma_k\varphi_k'+\Omega_k^2\varphi_k
=\mathcal J_{sss,k}+\mathcal J_{ss\gamma,k},
\label{eq:varphi-NL-finite}
\end{equation}
where the retained sources scale as
\begin{equation}
\begin{split}
\mathcal J_{sss,k}\sim\frac{k^6}{2K_k}\Bigg\{
&a\,c_1^{(0;3,0)}\beta_{0,k}
\left[(\beta_0\varphi+\beta_1\varphi')^2\right]_k
\\
&-\partial_\eta\left[
a\,c_1^{(0;3,0)}\beta_{1,k}
\left[(\beta_0\varphi+\beta_1\varphi')^2\right]_k
\right]
\Bigg\},
\end{split}
\label{eq:backreaction-scalar-source-phi}
\end{equation}
\begin{equation}
\begin{split}
\mathcal J_{ss\gamma,k}\sim\frac{k^4}{2K_k}\Bigg\{
&a\,c_1^{(0;3,0)}\beta_{0,k}
\left[\gamma'(\beta_0\varphi+\beta_1\varphi')\right]_k
\\
&-\partial_\eta\left[
a\,c_1^{(0;3,0)}\beta_{1,k}
\left[\gamma'(\beta_0\varphi+\beta_1\varphi')\right]_k
\right]
\Bigg\}.
\end{split}
\label{eq:backreaction-mixed-source-phi}
\end{equation}
The brackets denote momentum convolutions, with tensor
contractions suppressed and each internal coefficient evaluated
at the momentum of the field it multiplies. Time derivatives
act on the entire product. On the linear scalar solution,
Eq.~\eqref{eq:Phi-eom} gives
\begin{equation}
\begin{split}
\left(\beta_{0,k}\varphi_k+\beta_{1,k}\varphi_k'\right)'
={}&\left(\beta_{0,k}'-\Omega_k^2\beta_{1,k}\right)\varphi_k
\\
&+\left(\beta_{0,k}+\beta_{1,k}'
-\Gamma_k\beta_{1,k}\right)\varphi_k'.
\end{split}
\label{eq:backreaction-shift-derivative}
\end{equation}
Thus the sources involve only $\varphi$, $\varphi'$, and tensor
perturbations.

With the sources evaluated on the appropriate lower-order solutions, the local criterion for perturbative control is
\begin{equation}
\max\left\{
(\mathcal J_{sss,k})_{\rm rms},
(\mathcal J_{ss\gamma,k})_{\rm rms}
\right\}
\lesssim 
|\Gamma_k|(\varphi')_{\rm rms}
+|\Omega_k^2|\varphi_{\rm rms}.
\label{eq:backreaction-finite-bounds}
\end{equation}
The rms amplitudes are evaluated over a characteristic logarithmic
momentum interval, using oscillation envelopes where appropriate.
This comparison uses the magnitudes of the individual linear
terms and depends on the finite-$\sigma$ transfer functions.
Control of the accumulated response additionally requires the
sourced scalar correction to remain small over the interval of
interest.

\subsection{Large-$\lvert\sigma\rvert$ limit}

We take the large-$\lvert\sigma\rvert$ limit of the finite-$\sigma$
reduction at fixed $k$ and background time. At leading order,
\begin{equation}
K_k\longrightarrow K_\infty
=\frac{6a^2\mathcal H^2}{\bar\chi'^2},
\qquad
\Gamma_k\longrightarrow0,
\qquad
\Omega_k^2\longrightarrow\frac{k^2}{3}.
\end{equation}
The corresponding quadratic action is
\begin{equation}
S_{\varphi,\infty}^{(2)}
=\int d\eta\,\frac{d^3k}{(2\pi)^3}
\frac{6a^2\mathcal H^2}{\bar\chi'^2}
\left(
\varphi_k'\varphi_{-k}'
-\frac{k^2}{3}\varphi_k\varphi_{-k}
\right).
\label{eq:backreaction-S2-large}
\end{equation}
Since $a^2\mathcal H^2$ and $\bar\chi'$ are constant during
radiation domination, variation of the quadratic action gives
\begin{equation}
\frac{\delta S_{\varphi,\infty}^{(2)}}
{\delta\varphi_{-k}(\eta)}
=-\frac{12a^2\mathcal H^2}{\bar\chi'^2}
\left(
\varphi_k''+\frac{k^2}{3}\varphi_k
\right).
\label{eq:backreaction-S2-variation-large}
\end{equation}

We further restrict the estimate to comparable subhorizon
momenta, $k\gg\mathcal H$. Using
$\varphi_k'\sim k\varphi_k$ at the level of oscillation envelopes,
the shift constraint reduces to
\begin{equation}
B_k\simeq
\frac{6\mathcal H^2}{\bar\chi'k^2}\varphi_k,
\qquad
\partial_i\partial_jB
\sim\frac{\mathcal H^2}{\bar\chi'}\varphi.
\label{eq:B-subhorizon}
\end{equation}
The contribution proportional to $\varphi_k'$ in the shift
solution is suppressed by $(\mathcal H/k)^3$. We assume that
the coupling varies on time scales no shorter than the
characteristic oscillation time. For oscillating scalar and tensor modes with comparable wavenumbers, time derivatives in the cubic variation do not compensate for this suppression.
Suppressing numerical coefficients and angular kernels, the
retained cubic effective interactions scale as
\begin{equation}
S_{c_1,\mathrm{eff},\infty}^{(3)}
\sim\int d\eta\,d^3x\,
a\,c_1^{(0;3,0)}
\left[
\frac{\mathcal H^6}{\bar\chi'^3}\varphi^3
+\frac{\mathcal H^4}{\bar\chi'^2}\gamma'\varphi^2
\right].
\label{eq:backreaction-S3-scaling}
\end{equation}
The products in this expression represent schematic
contractions with dimensionless spatial operators.
Varying the retained cubic effective action and dividing by
$2K_\infty=12a^2\mathcal H^2/\bar\chi'^2$ gives the parametric
scalar equation
\begin{equation}
\begin{split}
\varphi_k''+\frac{k^2}{3}\varphi_k
&\sim\frac{\bar\chi'^2}{12a^2\mathcal H^2}
a\,c_1^{(0;3,0)}
\left[
\frac{\mathcal H^6}{\bar\chi'^3}(\varphi^2)_k
+\frac{\mathcal H^4}{\bar\chi'^2}(\gamma'\varphi)_k
\right]
\\
&\sim\frac{c_1^{(0;3,0)}}{a}
\left[
\frac{\mathcal H^4}{\bar\chi'}(\varphi^2)_k
+\mathcal H^2(\gamma'\varphi)_k
\right].
\end{split}
\label{eq:varphi-NL-final}
\end{equation}
Here the products denote momentum convolutions, with numerical
coefficients and dimensionless angular kernels suppressed.

Comparing the source amplitudes with the characteristic linear
terms $k^2\varphi_{\rm rms}$ gives the local estimates
\begin{equation}
\frac{|c_1^{(0;3,0)}|}{a}
\frac{\mathcal H^4}{|\bar\chi'|k^2}
\varphi_{\rm rms}\lesssim 1,
\qquad
\frac{|c_1^{(0;3,0)}|}{a}
\frac{\mathcal H^2}{k^2}
\gamma'_{\rm rms}\lesssim 1.
\label{eq:backreaction-parameters}
\end{equation}
The amplitudes refer to root-mean-square fluctuations over a
characteristic logarithmic momentum interval and, for the scaling
estimate, to their oscillation envelopes.

The regular scalar solution and the primordial curvature
normalization imply
\begin{equation}
\varphi_k(\eta)
=-\frac{\sqrt3\,\bar\chi'}{k}
\zeta_k^{\rm p}
\sin\left(\frac{k\eta}{\sqrt3}\right),
\qquad
\varphi_{\rm rms}
\sim\frac{|\bar\chi'|}{k}\sqrt{\mathcal P_\zeta(k)}.
\end{equation}
For oscillatory subhorizon tensor modes,
$\gamma'_{\rm rms}\sim k\sqrt{\mathcal P_\gamma(k,\eta)}$,
where the tensor spectrum is evaluated at the time of interest.
The local conditions become
\begin{equation}
\left|\frac{c_1^{(0;3,0)}k}{a}\right|
\frac{\sqrt{\mathcal P_\zeta(k)}}{x^4}\lesssim 1,
\qquad
\left|\frac{c_1^{(0;3,0)}k}{a}\right|
\frac{\sqrt{\mathcal P_\gamma(k,\eta)}}{x^2}\lesssim 1.
\label{eq:backreaction-spectral}
\end{equation}
For freely propagating tensor modes during radiation domination,
the amplitude decays as $x^{-1}$; this decay is already included
in $\mathcal P_\gamma(k,\eta)$.

\end{document}